\documentclass[amsmath,amssymb,aps,pra,twocolumn,
superscriptaddress,
]{revtex4-2}
\usepackage[utf8]{inputenc}
\usepackage[english]{babel}
\usepackage[T1]{fontenc}
\usepackage{amsmath,amssymb}
\usepackage{amsthm}
\usepackage{graphicx}      
\usepackage{wrapfig}

\usepackage{hyperref}
\usepackage{relsize}
\usepackage{tcolorbox}
\usepackage{thm-restate}
\usepackage[caption=false]{subfig}
\hypersetup{
    colorlinks=true,
    linkcolor=blue,
    citecolor=magenta,
    filecolor=magenta,      
    urlcolor=blue,
    pdftitle={},
    pdfpagemode=FullScreen,
    }
\usepackage{soul}
\usepackage{mathtools}

\usepackage{tikz}
\usetikzlibrary{quantikz}
\usetikzlibrary{shapes}
\usepackage{xcolor}
\theoremstyle{definition}
\usepackage{float}
\usepackage{makecell}

\makeatletter
\newlength\min@xx

\makeatother

\usepackage{lipsum}
\usepackage{algorithm}
\usepackage[noend]{algpseudocode}

\usepackage{physics}

\begin{document}

\title{The Cost of Entanglement Renormalization on a Fault-Tolerant Quantum Computer}
\date{\today}

\author{Joshua Job}
\affiliation{Lockheed Martin, Sunnyvale, CA, 94089}
\author{Isaac H. Kim}
\affiliation{PsiQuantum, Palo Alto, CA, 94304 }
\author{Eric Johnston}
\affiliation{PsiQuantum, Palo Alto, CA, 94304 }
\author{Steve Adachi}
\affiliation{Lockheed Martin, Sunnyvale, CA, 94089}

\begin{abstract}
  We perform a detailed resource estimate for the prospect of using deep entanglement renormalization ansatz (DMERA) on a fault-tolerant quantum computer, focusing on the regime in which the target system is large. For probing a relatively large system size ($64\times 64$), we observe up to an order of magnitude reduction in the number of qubits, compared to the approaches based on quantum phase estimation (QPE). We discuss two complementary strategies to measure the energy. The first approach is based on a random sampling of the local terms of the Hamiltonian, requiring $\mathcal{O}(1/\epsilon^2)$ invocations of quantum circuits, each of which have depth of at most $\mathcal{O}(\log N)$, where $\epsilon$ is the relative precision in the energy and $N$ is the system size. The second approach is based on a coherent estimation of the expectation value of observables averaged over space, which achieves the Heisenberg scaling while incurring only a logarithmic cost in the system size. For estimating the energy per site of $\epsilon$,  $\mathcal{O}\left(\frac{\log N}{\epsilon} \right)$ $T$ gates and $\mathcal{O}\left(\log N \right)$ qubits suffice. The constant factor of the leading contribution is shown to be determined by the depth of the DMERA circuit, the gates used in the ansatz, and the periodicity of the circuit. We also derive tight bounds on the variance of the energy gradient, assuming the gates are random Pauli rotations. 
\end{abstract}
\maketitle

\section{Introduction}
\label{sec:intro}

One of the most impactful applications of quantum computers is expected to be the simulation of realistic materials that appear in nature~\cite{McArdle2020}. This idea, which goes back to the seminal work of Feynman~\cite{Feynman1982} and Lloyd~\cite{Lloyd1996}, has recently gained much interest with the rapid advancement in quantum computing technology.

Many of the applications considered in the literature had been focused on simulation of a small molecule associated with the nitrogen fixation process, known as ``FeMoco'' in the literature. Reiher et al.~\cite{Reiher2017} was the first to perform a rigorous resource estimate for this system. Since then, improvements were made in the efficiency of this algorithm by adopting the block-encoding approach of Low and Chuang~\cite{Low2019} and by using cleverly designed ways to block-encode the Hamiltonian~\cite{Berry2019,Lee2021,vonBurg2021}.

More recently, quantum algorithms for simulating material properties of a solid state system have been studied~\cite{Babbush2018,Babbush2019,Su2021,Delgado2022,Zini2023}. The primary challenge in this direction is that the number of orbitals one needs to consider is much larger than that required for simulating small molecules. While there are many motivating examples, let us consider the case of multiferroic materials~\cite{Fiebit2016,Spaldin2019} as a concrete example. These materials, which exhibit a coupling between their magnetic (magnetization) and electric (polarization) properties, have been a subject of intense experimental and theoretical interest over the past few decades. A prototypical example of a multiferroic is an epitaxially grown $\text{BiFeO}_3$ thin film~\cite{Wang2003}. These materials consist of domains, exhibiting either electric or antiferromagnetic order~\cite{Catalan2009} and the bulk properties of the material depend on the sizes of these domains. The typical side of the domain consists of $100\times 100$ unit cells if not larger~\cite{Chu2007}. Therefore, to accurately simulate realistic properties of this material, one would need to incorporate at least tens of thousands of orbitals, at least two orders of magnitude larger than the number required for small molecules~\cite{Reiher2017,Berry2019,Lee2021,vonBurg2021}. 

These new class of algorithms~\cite{Babbush2018,Babbush2019,Su2021,Delgado2022,Zini2023} --- known as first quantization methods --- have better asymptotic scaling for the number of qubits and the $T$-gate/Toffoli count compared to methods in Ref.~\cite{Berry2019,Lee2021,vonBurg2021}. However, careful resource estimates in this direction led to costs which are still a few orders of magnitude larger than what is required for Ref.~\cite{Berry2019,Lee2021,vonBurg2021}. Therefore, in order to use these methods, one would likely need a larger fault-tolerant quantum computer. 

\begin{comment}
  However, as noted in Ref.~\cite{Lee2021}, it might be difficult to make further improvements in the asymptotic complexity of these algorithms for arbitrary basis sets. The state-of-the-art method~\cite{Lee2021} achieves a near-linear scaling in the size of the basis set, norm of the Hamiltonian, and inverse precision, which are all likely optimal. If that is the case, possible improvements can be only made by choosing a more structured basis set, such as the dual of the plane wave basis~\cite{Babbush2018}. However, such basis sets tend to require a larger number of orbitals to yield the same degree of accuracy. Since the number of qubits needed grows at least linearly with the number of orbitals in any approach based on quantum phase estimation, possible improvements in the $T$-gate cost may come with a price: a large number of logical qubits.
\end{comment}

The purpose of this paper is to explore an alternative approach to study properties of materials on a quantum computer. The approach we investigate will be variational in nature~\cite{Peruzzo2013}, employing the deep entanglement renormalization ansatz (DMERA)~\cite{Kim2017}, a variant of Vidal's multi-scale entanglement renormalization ansatz~\cite{Vidal2008}. While there may be other ansatzes that one may choose for this purpose, we investigated DMERA primarily for two reasons. First of all, it is well-known that, for the purpose of measuring expectation values of local observables, the DMERA circuit can be compressed to a circuit applied to a constant number of qubits, independent of the size of the system~\cite{Vidal2008,Evenbly2011,Kim2017}. The accuracy of the computation increases progressively as the number of qubits being used increases. Therefore, even with a small quantum computer, one can hope to approximate ground state properties of a much larger system. Secondly, the DMERA ansatz is known to be suitable to describe a large class of phases of matter, such as topologically ordered systems~\cite{Aguado2008,Konig2009} and systems near the point of quantum phase transition~\cite{Vidal2008,Haegeman2017,Evenbly2016}. These studies suggest that DMERA is an ansatz capable of approximating a wide class of physical systems. 

The main advantage of approaches like DMERA is that it leverages the existing progress made in the tensor network literature~\cite{White1992,Verstraete2006,Vidal2008}; see Ref.~\cite{TNReview} for a review. These are classical computational methods whose underlying complexity is quantified in terms of the ``bond dimension,'' related to the number of parameters in the ansatz. Often times a better approximation to the ground state properties can be obtained by increasing the bond dimension. It was realized in Ref.~\cite{Kim2017,Kim2017a,Kim2017c} that the tensor network approach can be employed on a quantum computer, used as an ansatz for the variational quantum eigensolver (VQE)~\cite{Peruzzo2013}. The key observation was that the number of qubits and gates needed on the quantum computer needed to compute physical properties of interest, e.g., correlation functions and energy, scales \emph{logarithmically} with the bond dimension. The practical bottleneck in tensor network approaches has been an unfavorable scaling in the bond dimension. Since a quantum computer can improve this scaling substantially, the tensor network approach can be potentially sped up substantially by a quantum computer.

In the case of DMERA, the speedup one can hope to obtain can be quantified in the following way. The ``size'' of the DMERA ansatz can be quantified in terms of two parameters, $D$ and $n$, which are related to the total depth of the circuit and the width of the circuit as $Dn$ and $2^{dn}$, respectively. (Here $d$ is the number of spatial dimensions.) A quantum computer can compute expectation values of local observables in $\mathcal{O}(Dn)$ time using $\mathcal{O}(D^d)$ qubits. In contrast, a classical computer requires memory and computation time that scale \emph{exponentially} with $D^d$. The dependence on $n$ is the same in both cases, but the exponential dependence on $D^d$ makes a classical simulation challenging. Therefore, one can expect a substantial speedup in estimating the expectation values of local observables. 

In spite of these advantages, there are obvious concerns one can raise for this approach. Existing studies in variational quantum algorithms have shown that there are several potential issues one must be wary about. The first issue is that often the sheer number of measurements needed to estimate the energy can be quite large in practice~\cite{Wecker2015}. Another important issue is that the gradient tends to vanish exponentially with the circuit depth~\cite{McClean2018}. Understanding whether these issues persist in DMERA shall be the primary goals of this paper. 

To that end, we carry out a detailed study on the prospect of using DMERA for interacting fermions in two spatial dimensions, focusing on the cost needed for a fault-tolerant quantum computer. We do so by first critically analyzing the issue of barren plateaus~\cite{McClean2018}. ``Barren plateau'' refers to a phenomenon in which the gradient of a parameterized quantum circuit decays exponentially with the circuit depth. We compute the typical value of the gradient and observe that its magnitude decays much slower than what was suggested in structureless ansatzes. Specifically, we find that the variance depends greatly on the location of the gates. For instance, the gates that appear towards the end of the circuit has a variance decaying exponentially \emph{only} in $D$, but not in $n$. The gates that appear early, on the other hand, does decay exponentially in both $D$ and $n$. This is similar to the conclusion recently made in Ref.~\cite{Barthel2023absence}. Though the assumptions we make on the gates are different, our conclusions agree. 

Secondly, we perform a detailed resource estimate for estimating a variational energy up to a fixed precision. We obtain a bound sharper than the one derived in Ref.~\cite{Kim2017c} and calculate the exact numbers numerically. The resource estimate was performed for two different schemes for measuring the energy, which are complementary to each other. We refer to the first approach as \emph{incoherent sampling}, which is based on a sampling of the expectation values of the local terms in the Hamiltonian. An encouraging fact about this approach is that the number of qubits needed is substantially lower than the approaches based on QPE.

However, because this approach has a quadratic scaling in the inverse precision, the number of gates needed is greater compared to QPE-based approaches. We have estimated these numbers in detail for a concrete physical system: the Fermi-Hubbard model in two spatial dimensions. We chose a relatively large system size of $64\times 64$, as a proxy for understanding the performance of this method applied to the case of studying $\text{BiFeO}_3$ film. Comparing the gate count estimate for the same model using the state-of-the-art approach based on QPE~\cite{Kivlichan2020,Campbell2021}, we find the incoherent approach to require a larger number of $T$ gates. Specifically, we find the number of $T$-gates needed is likely at least an order of magnitude larger than that required for QPE. Therefore, while the incoherent approach may be more advantageous for a small fault-tolerant quantum computer, for a large enough quantum computer it is expected to be inferior compared to the conventional QPE-based approaches.

Fortunately, Heisenberg scaling can be restored under a reasonable assumption on the choice of gates used in the DMERA ansatz. We show that, if the gates chosen in the DMERA circuit is periodic with some finite period, one can exploit this to improve the complexity of estimating the variational ground state energy. Specifically, we improve the dependence on inverse precision from $\mathcal{O}(1/\epsilon^2)$ to $\mathcal{O}(1/\epsilon)$ whilst only increasing the qubit count from $\mathcal{O}(1)$ to $\mathcal{O}(\log N)$, where $N$ is the system size. Since such periodicity is present in many practical calculations~\cite{Evenbly2011,Evenbly2016}, we expect this assumption to be a realistic one. This is a novel amalgamation of VQE~\cite{Peruzzo2013} and amplitude estimation~\cite{Brassard2002}, which may be useful in a broader context.

While carrying out these resource estimates, we obtain the following additional results, which may be of independent interest. We introduce a novel compilation technique that can reduce the number of qubits substantially. We observe up to two-fold reduction in the number of qubits for DMERA [Fig.~\ref{fig:dmera_incoherent_cost}]. While our main motivation is to reduce the qubit count for DMERA, the algorithm itself is applicable to any quantum circuits. Moreover, we introduce novel compilation techniques for a fermionic variant of DMERA. This is an ansatz defined in terms of a fermionic gates. A similar ansatz was already introduced and studied in Ref.~\cite{Corboz2009}. In particular, it is well-known that the efficient contractibility of MERA remains intact for this extension. The fact that this efficient contraction algorithm can be ported to a quantum circuit (with a substantially lower complexity) can be guaranteed on general grounds~\cite{Arad2010}. However, we provide a more explicit and optimized algorithm which achieves the same goal. Our novel contribution is an explicit realization of a physical operation (in terms of one- and two-qubit gates) that ``resets'' the fermion occupation number, using the Jordan-Wigner transformation~\cite{JW1928}. Using this technique, we were able to extend our resource estimate for qubits to fermions. 

The rest of the paper is structured as follows. In Section~\ref{section:dmera}, we review DMERA~\cite{Vidal2008,Kim2017}. In Section~\ref{sec:pcc} we review the concept of past causal cone and elucidate its relationships with the parameters that quantify the size of a DMERA ansatz. In Section~\ref{sec:barren_plateau}, we estimate the typical value of the gradient in DMERA, quantifying it in terms of the parameters that determine the size of DMERA. In particular, we show that the gradient decays much slower than what one would naively expect, sidestepping the barren plateau problem~\cite{McClean2018}. In Section~\ref{sec:resource_estimate_incoherent}, we perform a resource estimate for the number of gates and qubits. In Section~\ref{sec:amplitude_estimation}, we describe a quantum algorithm that improves the asymptotic cost of the algorithm in Section~\ref{sec:resource_estimate_incoherent} in terms of gate count, with a logarithmic increase in the qubit count. We conclude with a set of open problems in Section~\ref{sec:conclusion}.

\section{DMERA \label{section:dmera}}

Let us begin by introducing the central piece of our work: \emph{deep multi-scale entanglement enormalization ansatz}, or DMERA for short~\cite{Kim2017c}. DMERA is a variant of the well-known multi-scale entanglement renormalization ansatz (MERA)~\cite{Vidal2008}, which is a variational ansatz that has been used successfully to study ground state properties of quantum many-body systems at and away from quantum critical points~\cite{Vidal2008,Evenbly2009}. The key advantage of MERA, compared to other variational ansatzes, is that it can produce a power-law decay of correlation, a feature that is commonly present in quantum many-body systems at their quantum critical points. 

If we can study the quantum critical points of a general locally interacting Hamiltonian, we would be able to obtain a wealth of information. Quantum critical points often contain a wide variety of universal data that characterizes the underlying phase, such as the critical exponent and the central charge of the underlying theory. Such data can often pinpoint the underlying theory to a select few if not one, which can then constrain the phases that lie on both sides of the transition in a highly nontrivial way~\cite{Grover2014}. As such, a successful application of MERA to higher dimensions can potentially lead to tremendous breakthroughs in understanding exotic phenomena that can occur in interacting quantum many-body systems, such as high-temperature superconductivity~\cite{Bednorz1986}.

Unfortunately, while MERA has been used successfully in one spatial dimension (1D)~\cite{Vidal2008,Evenbly2009}, its usage in two spatial dimensions (2D)~\cite{Cincio2008,Evenbly2009,Evenbly2010} and above has been relatively limited. This is mainly due to the fact that the computational cost of using MERA can be prohibitive in practice. Specifically, suppose we want to study a system consisting of $N$ particles and we are interested in estimating the computational cost of measuring an expectation value of a local observable. This cost scales merely \emph{logarithmically} with $N$~\cite{Evenbly2009}, but the ``proportionality constant'' in front of the logarithm is too enormous in practice. 

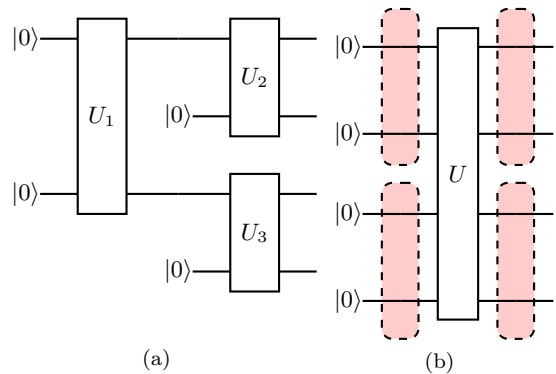
\begin{figure}[h]
    \centering
    \subfloat[]{
    \begin{quantikz}
    \ket{0}& \gate[3, nwires={2}]{U_1} & \qw &\gate[2]{U_2}&\qw  \\
    & &\ket{0} & &\qw\\
    \ket{0}& & \qw &\gate[2]{U_3}&\qw\\
    & &\ket{0} & &\qw \\
    \end{quantikz}
    }
    \subfloat[]{
    \begin{quantikz}[row sep=0.75cm]
    \ket{0} &\qw\gategroup[2,steps=1,style={dashed,
                   rounded corners,fill=red!20, inner xsep=2pt},
                   background]{} & \gate[4]{U} & \qw\gategroup[2,steps=1,style={dashed,
                   rounded corners,fill=red!20, inner xsep=2pt},
                   background]{} &\qw \\
    \ket{0} &\qw & & \qw & \qw \\
    \ket{0} &\qw\gategroup[2,steps=1,style={dashed,
                   rounded corners,fill=red!20, inner xsep=2pt},
                   background]{} & & \qw\gategroup[2,steps=1,style={dashed,
                   rounded corners,fill=red!20, inner xsep=2pt},
                   background]{} & \qw \\
    \ket{0} &\qw & & \qw &\qw
    \end{quantikz}
    }
    \caption{(a) 1D MERA with bond dimension $\chi=2$ over $4$ qubits. (b) An arbitrary unitary acting on $4$ qubits. Formally, this can be viewed as a 1D MERA with bond dimension $\chi=4$ over $4$ qubits by viewing $U$ as a unitary acting on two copies of $\chi^2=4$-dimensional Hilbert spaces (red).}
    \label{fig:dmera_simple}
\end{figure}

To understand where this large constant comes from, it is helpful to consider a simple toy example of MERA, which we describe in Figs.~\ref{fig:dmera_simple}. As one can see in these figures, MERA is simply a quantum circuit applied to a fixed product state such as $|0\ldots 0\rangle$. Each unitary in Fig.~\ref{fig:dmera_simple} generally acts on a $(\chi \times \chi)$-dimensional Hilbert space, where $\chi=2$ and $4$ in Fig.~\ref{fig:dmera_simple}(a) and (b) respectively. Assuming that the unitaries in these circuit diagrams can take any value in $U(\chi^2)$, the ansatz in Fig.~\ref{fig:dmera_simple}(a) forms a strict subfamily of Fig.~\ref{fig:dmera_simple}(b). Therefore, a smaller bond dimension equates to studying a smaller family of quantum states.  

This parameter $\chi$ -- also known as the \emph{bond dimension} -- is precisely the one that determines the proportionality constant in front of $\log N$. For 1D MERA, using the state-of-the-art method, the dependence can grow as $\mathcal{O}(\chi^5)$~\cite{Evenbly2015}. In 2D MERA, depending on the geometry of the quantum circuit, one gets anywhere between $\mathcal{O}(\chi^{16})$ to $\mathcal{O}(\chi^{24})$~\cite{Evenbly2009,Cincio2008,Evenbly2010}. While the computational cost of MERA is ``polynomial'' in the bond dimension, the degree of the polynomial is so large that it is very difficult to make it practical. To obtain a more accurate approximation of ground states, $\chi$ must be increased. However, this enormous degree is what prevents us from using a large value of $\chi$

Fortunately, a near-term quantum computer can potentially come to the rescue to alleviate this challenging problem~\cite{Kim2017c}. Here, the computational cost merely grows \emph{logarithmically} with $\chi$, and as such, the enormous constant prefactor can be significantly reduced on a quantum computer.

Below, we will provide a systematic approach to constructing DMERA. Roughly speaking, DMERA can be thought of as a variant of MERA in which the unitaries acting on $(\chi\times \chi)$-dimensional Hilbert space are broken down into a sequence of two-qubit gates acting on $\lceil \log_2(\chi) \rceil$ qubits.

\begin{figure*}[t]
    \centering
    \subfloat[][]{
    \begin{quantikz}[row sep=0.75cm]
    \lstick[wires=5]{$\ket{\psi}$} &\qw\gategroup[6,steps=2,style={dashed,
                   rounded corners,fill=green!20, inner xsep=2pt},
                   background]{{\sc Fine-graining}} & \qw &\qw  \\
    &\ket{0}& \qw  &\qw\\
    &\qw &  \qw  &\qw\\
    &\ket{0}& \qw  &\qw \\
    &\qw &  \qw  &\qw \\
    &\ket{0}& \qw   &\qw
    \end{quantikz}
    }
     \subfloat[][]{
    \begin{quantikz}[row sep=0.35cm]
    & \gate[2]{U_1} \gategroup[6,steps=3,style={dashed,
                   rounded corners,fill=blue!20, inner xsep=2pt},
                   background]{{\sc Finite-depth quantum circuit}}& \qw& \gate[2]{U_6} &\qw\\
    &  &\gate[2]{U_4} & &\qw\\
    & \gate[2]{U_2}& & \gate[2]{U_7}&\qw\\
    & &\gate[2]{U_5} & &\qw\\
    & \gate[2]{U_3}& &\gate[2]{U_8}&\qw\\
    & &\qw& &\qw
    \end{quantikz}
    }
    \subfloat[][]{
    \begin{quantikz}[row sep=0.6cm]
    \ket{0}&\qw\gategroup[5,steps=2,style={
                   rounded corners,fill=blue!20, inner xsep=2pt}]{}\qw &\qw  &\qw \slice{1} &\qw & \qw\gategroup[7,steps=2,style={
                   rounded corners,fill=green!20, inner xsep=2pt}]{} & \qw &\qw &\qw  \gategroup[7,steps=2,style={
                   rounded corners,fill=blue!20, inner xsep=2pt}]{} &\qw &\qw\slice{2} &\qw &\qw  \gategroup[8,steps=2,style={
                   rounded corners,fill=green!20, inner xsep=2pt}]{} &\qw &\qw &\qw \gategroup[8,steps=2,style={
                   rounded corners,fill=blue!20, inner xsep=2pt}]{} &\qw &\qw \slice{3}&\qw \\
    & & & &&&  & & & & & & &\qw &\qw &\qw &\qw &\qw&\qw\\
    & & & && & \qw &\qw &\qw &\qw  &\qw &\qw &\qw &\qw &\qw &\qw &\qw &\qw&\qw\\
    & & & &  & & & & & & & & &\qw &\qw &\qw &\qw &\qw&\qw\\
    \ket{0}&\qw &\qw&\qw &\qw & \qw& \qw & \qw &\qw \qw &\qw &\qw &\qw &\qw &\qw &\qw &\qw &\qw &\qw&\qw\\
    & & & & & & & & & & & &  &\qw &\qw &\qw &\qw &\qw&\qw\\
    & & & & & & \qw &\qw &\qw & \qw  &\qw &\qw &\qw &\qw &\qw &\qw &\qw &\qw&\qw\\
    & & & & & & & & & & & & &\qw &\qw &\qw &\qw &\qw&\qw
    \end{quantikz}
    }
    \caption{(a) Fine-graining a three-qubit state $\langle \psi|$ to $\langle \psi| \langle 000|$ (b) A depth-$3$ quantum circuit (c) DMERA as an alternating sequence of finite-depth quantum circuit and fine-graining.}
    \label{fig:dmera_def}
\end{figure*}
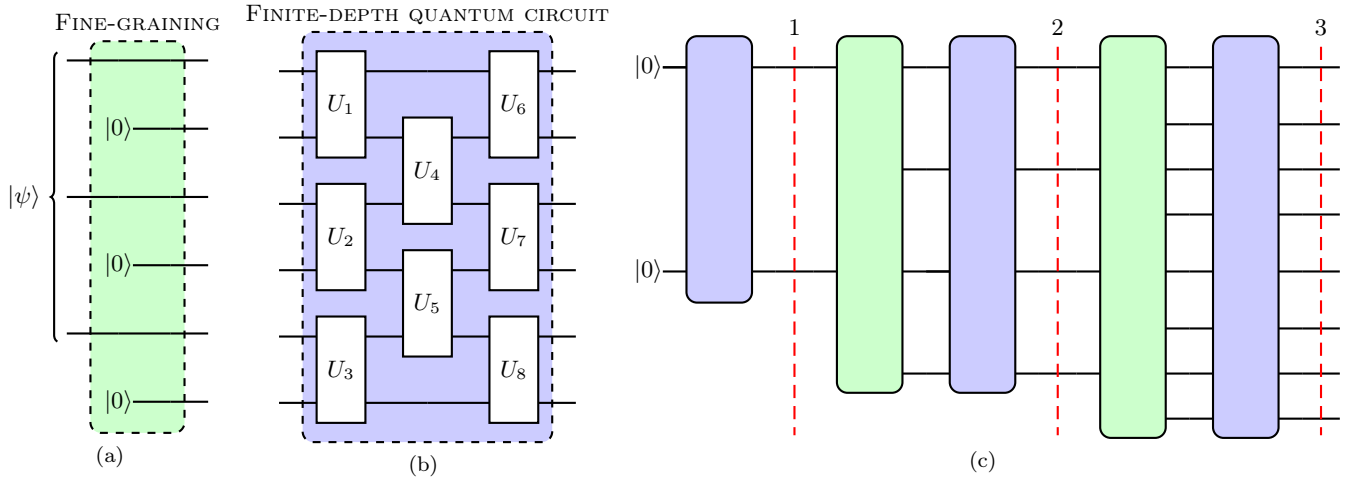

More formally, we can define DMERA as follows. Instead of explicitly drawing the circuit, we will define DMERA in terms of two abstract procedures: \emph{fine-graining} and \emph{finite-depth quantum circuit}. Mathematically, fine-graining refers to a process in which a general quantum state $|\psi\rangle$ is embedded into a larger Hilbert space by appending a simple state such as $|0\ldots 0\rangle$.  Finite-depth quantum circuit means a unitary consisting of geometrically local gates with depth that is independent of $N$. DMERA is then defined as an alternating sequence of finite-depth quantum circuit and fine-graining [Fig.~\ref{fig:dmera_def}].

Note that there is a great deal of freedom in choosing the fine-graining procedure. The number of newly introduced qubits, as well as the way in which the qubits are arranged with respect to another, are the important choices that can affect the performance of DMERA. Likewise, there are different choices one can make about finite-depth quantum circuits, such as the underlying gate set and where the gates act within each layer of the circuits.

To be quantitative in our statements, we will have to make some choices, which we describe as follows\footnote{Changing these choices will not alter the conclusions of this paper significantly, although the precise numbers we estimate later are likely to depend on those choices.}. We shall focus on physical systems in which the qubits are arranged on a (hyper-)cubic lattice of dimension $d$. We will only consider $(r_1,\ldots, r_d)$-\emph{regular} fine-graining procedures, with $r_i=2$, meaning that the qubits appearing prior to the fine-graining procedure are embedded evenly in a larger lattice elongated in the $i$'th direction by a factor of $r_i=2$ for $i=1,\ldots, d$. We shall refer to these constants as \emph{expansion coefficients}. As for the finite-depth unitaries, we shall decompose them into unit-depth quantum circuits, each of which consists of non-overlapping two-qubit gates acting on nearest-neighbors in a fixed direction. For instance, if we fix the direction to $\hat{x}$, we can consider a tensor product of unitaries acting on  $(2i, \ldots)$ and $(2i+1, \ldots)$ for $i\in \mathbb{Z}$, where the rest of the coordinates are the same and can take an arbitrary integer value within the hypercube considered. Alternatively, one can consider a tensor product of unitaries acting on $(2i-1, \ldots)$ and $(2i, \ldots)$. 

To summarize, DMERA is defined in terms of a sequence of fine-graining procedures and finite-depth quantum circuits. We can formally write this as
\begin{equation}
  |\Psi\rangle = \mathcal{U}_n \mathcal{F}_N \ldots \mathcal{U}_1 \mathcal{F}_1,
\end{equation}
where the $\mathcal{F}_i$ and $\mathcal{U}_i$ are the $i$'th fine-graining procedure and finite-depth quantum circuit. Here $n$ is a parameter that quantifies the size of the system. Specifically, let $L$ be the linear size of the system. Assuming that the DMERA is $r$-regular, the linear size is related to $n$ as $L= r^n$. Thus, we will focus on understanding the computational cost of using DMERA in terms of the following three parameters.
\begin{equation*}
\boxed{
    \begin{aligned}
        n: &\text{ Number of scales} \\
        r: &\text{ Expansion coefficient} \\
        D: &\text{ Circuit depth of each finite-}\\
         &\text{ depth quantum circuit}
    \end{aligned}
    }
\end{equation*}
To be more precise, we will assume that the underlying DMERA is $r$-regular and moreover assume that every $\mathcal{U}_i$ has a circuit depth of $D$ with a sequence of unit-depth quantum circuits described in the paragraph above.

\section{Past causal cone}
\label{sec:pcc}

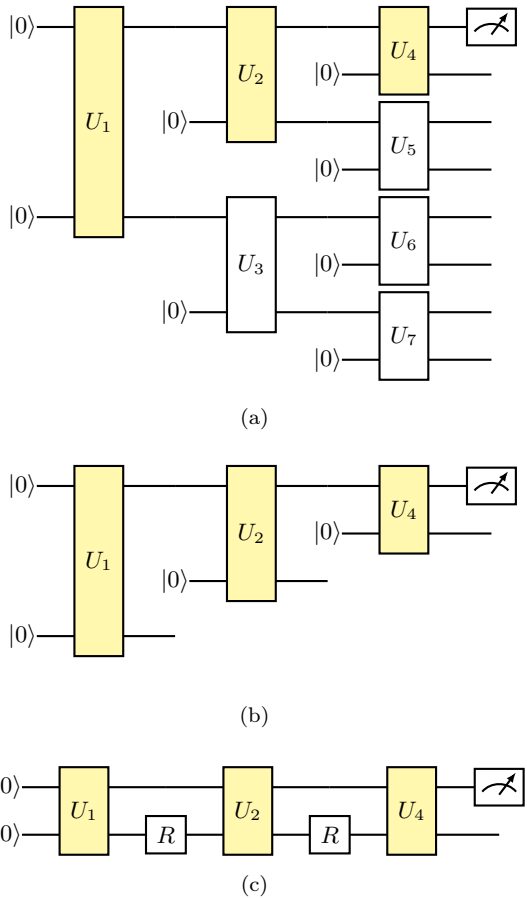
\begin{figure}[h]
    \centering
    \subfloat[]{
    \begin{quantikz}[row sep=0.1cm]
    \ket{0}& \gate[5, nwires={2,3,4}, style={fill=yellow!40}]{U_1} &\qw &\gate[3, nwires={2}, style={fill=yellow!40}]{U_2} &\qw &\gate[2, style={fill=yellow!40}]{U_4}&\meter{}\\
    & & & &\ket{0} & &\qw\\
    & &\ket{0} & &\qw &\gate[2]{U_5}&\qw\\
    & & & &\ket{0}&&\qw\\
    \ket{0}& &\qw &\gate[3,nwires={2}]{U_3} &\qw&\gate[2]{U_6}&\qw\\
    & & & &\ket{0}&&\qw\\
    & &\ket{0} & &\qw&\gate[2]{U_7}&\qw\\
    & & & &\ket{0}&&\qw\\
    \end{quantikz}
    }
    \\
    \subfloat[]{
    \begin{quantikz}[row sep=0.1cm]
    \ket{0}& \gate[5, nwires={2,3,4}, style={fill=yellow!40}]{U_1} &\qw &\gate[3, nwires={2}, style={fill=yellow!40}]{U_2} &\qw &\gate[2, style={fill=yellow!40}]{U_4}&\meter{}\\
    & & & &\ket{0} & &\qw\\
    & &\ket{0} & &\qw &&\\
    & & & &&&\\
    \ket{0}& &\qw & &&&\\
    & & & &&&\\
    & & & &&&\\
    & & & &&&\\
    \end{quantikz}
    }
    \\
    \subfloat[]{
    \begin{quantikz}[row sep=0.1cm]
    \ket{0}& \gate[2, style={fill=yellow!40}]{U_1} &\qw &\gate[2, style={fill=yellow!40}]{U_2} &\qw &\gate[2, style={fill=yellow!40}]{U_4}&\meter{}\\
    \ket{0}& & \gate[]{R} & & \gate[]{R} & &\qw
    \end{quantikz}
    }
    \caption{(a) An example of a 1D DMERA with $n=3$ and $D=1$. Note that only the yellow gates can affect the measurement outcome of the top-most qubit. (b) After removing the irrelevant gates and qubits, we obtain the past causal cone. (c) By noting that some of the qubits are no longer in use, one can simply feed them into $|0\rangle$, leading to a width-reduced version of the past causal cone. Here $R$ resets the qubit to the $|0\rangle$ state.}
    \label{fig:past_causal_cone_dmera}
\end{figure}

While one can in principle directly implement the entire set of finite-depth quantum circuits and fine-graining procedures, that will not be the most economical choice. Instead, given an observable $\hat{O}$, one can simply discard all the gates that do not affect the expectation value of $\hat{O}$. What remains is the \emph{past causal cone} of $\hat{O}$~\cite{Giovannetti2008,Evenbly2009} [Fig.~\ref{fig:past_causal_cone_dmera}].

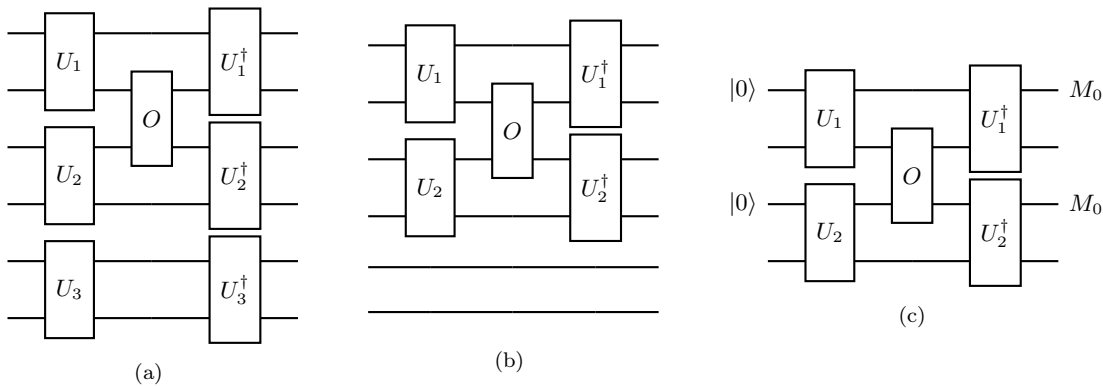
\begin{figure*}[t]
  \centering
     \subfloat[][]{
    \begin{quantikz}[row sep=0.1cm]
    & \gate[2]{U_1} & \qw& \gate[2]{U_1^{\dagger}} &\qw\\
    &  &\gate[2]{O} & &\qw\\
    & \gate[2]{U_2}& & \gate[2]{U_2^{\dagger}}&\qw\\
    & &\qw & &\qw\\
    & \gate[2]{U_3}& \qw &\gate[2]{U_3^{\dagger}}&\qw\\
    & &\qw& &\qw
    \end{quantikz}
     }
     \hspace{0.5cm}
     \subfloat[][]{
    \begin{quantikz}[row sep=0.1cm]
    & \gate[2]{U_1} & \qw& \gate[2]{U_1^{\dagger}} &\qw\\
    &  &\gate[2]{O} & &\qw\\
    & \gate[2]{U_2}& & \gate[2]{U_2^{\dagger}}&\qw\\
    & & \qw & &\qw\\
    & \qw &\ghost{X}\qw &\qw &\qw\\
    & \qw &\ghost{X}\qw &\qw &\qw
    \end{quantikz}
     }
     \hspace{0.5cm}
    \subfloat[][]{
    \begin{quantikz}[row sep=0.1cm]
    \lstick{$\ket{0}$} & \gate[2]{U_1} & \qw& \gate[2]{U_1^{\dagger}} & \rstick{$M_0$}\qw\\
    &  &\gate[2]{O} & &\qw\\
     \lstick{$\ket{0}$} & \gate[2]{U_2}& & \gate[2]{U_2^{\dagger}}& \rstick{$M_0$}\qw\\
    & & \qw & &\qw
    \end{quantikz}
     }
    \caption{(a) A unit-depth circuit applied to an observable in the Heisenberg picture. (b) Some of the unnecessary gates can be removed from (a), resulting in a reduced circuit.  (c) Fine-graining step in the Schr\"odinger picture becomes the coarse-graining step in the Heisenberg picture. Here $M_0$ is a post-selection of $|0\rangle$.}
    \label{fig:pcc_Heisenberrg}
\end{figure*}

In order to understand how to obtain the past causal cone, it is helpful to think in the Heisenberg picture. Specifically, recall that the DMERA wave function $|\Psi\rangle$ can be prepared by an alternating sequence of fine-graining and finite-depth quantum circuits, denoted as $(\mathcal{F}_i: i=1,\ldots, n)$ and $(\mathcal{U}_i, i=1,\ldots, n)$, respectively:
\begin{equation}
    |\Psi\rangle = \mathcal{U}_n\mathcal{F}_n\ldots \mathcal{U}_1\mathcal{F}_1.
\end{equation}
In particular, $\mathcal{F}_1 = |0\ldots 0\rangle$ is simply a preparation of a product state over $2^n$ qubits. In the Heisenberg picture, an observable ``evolves'' in the following way:

\begin{equation}
  \hat{O}_{k+1} =
    \mathcal{F}_{n-k}^{\dagger}\mathcal{U}_{n-k}^{\dagger} \hat{O}_{k} \mathcal{U}_{n-k} \mathcal{F}_{n-k}  k= 0\mod D.
  \label{eq:pcc_layer_by_layer}
\end{equation}
for $k \in \{0,\ldots, n-1 \}$. Given a support of $\hat{O}_k$, one can bound the support of $\hat{O}_{k+1}$ by removing the gates in $\mathcal{F}_{n-k}$ and $\mathcal{U}_{n-k}$ in the following way. To be more specific, note that $\mathcal{U}_{n-k}$ is a depth-$D$ quantum, consisting of $D$ layers of unit-depth quantum circuit. Within each unit-depth quantum circuit, we can remove the gates whose support is disjoint from the Heisenberg-evolved operator up to that point, since for any such $u$ it commutes with the operator. Reparing this $D$ times, we obtain a support that is expanded from that of $\hat{O}_k$ [Fig.~\ref{fig:pcc_Heisenberrg}(a), (b)]. Next, the fine-graining procedure in the Heisenberg picture becomes a \emph{coarse-graining} procedure, reducing the support of the observable [Fig.~\ref{fig:pcc_Heisenberrg}(c)]. 

Now we can bound the size of the support of $\hat{O}_k$ as follows. For simplicity, let us assume that $\hat{O}_k$ is supported on a ball of radius $R_k$. Assuming that the depth of $\hat{U}_{i}$ is bounded by $D$ for all $i$, the support of $\mathcal{U}_{n-k}^{\dagger}\hat{O}_k\mathcal{U}_{n-k}$ is bounded by $R_k+ D$. Assuming that the fine-graining procedure is $(r,\ldots, r)$-regular, after coarse-graining the radius is bounded by $\sim R_k/r + D/r$. Therefore, we obtain the following recursive inequality:
\begin{equation}
    R_{k+1} \lessapprox \frac{R_k}{r} + \frac{D}{r}.
\end{equation}
As $k\to \infty$, we can expect $R_{k+1}\approx R_k$, yielding
\begin{equation}
    R_k \approx \frac{D}{r-1}.
\end{equation}
Upon evolving the operator in the Heisenberg picture, the radius is again expanded at most by an additive factor of $D$, yielding the radius of the past causal cone of $Dr/(r-1)$. Thus, the size of the support of $\hat{O}_k$ remains as $((Dr/(r-1))^d$ for all $k$. The gates that map $\hat{O}_k$ to $\hat{O}_{k+1}$ are also all supported on a ball of radius $\mathcal{O}(Dr/(r-1))$. 

By converting this reduced circuit back into the Schr\"odinger picture, we finally obtain the past causal cone of the observable $\hat{O}$ [Fig.~\ref{fig:past_causal_cone_dmera}(b)].  Moreover, by noting that some of the qubits are no longer in use, one can reduce the width of the circuit to be independent of $n$ [Fig.~\ref{fig:past_causal_cone_dmera}(c)]. Thus, the circuit depth and the width of the past causal cone of DMERA for every local observable is
\begin{equation}\label{eq:pcc_width}
\boxed{
    \begin{aligned}
        \text{Depth} &= \mathcal{O}(nD), \\
        \text{Width} &= \mathcal{O}((Dr/(r-1))^d).
    \end{aligned}
}
\end{equation}
This entire process amounts to converting $\mathcal{U}_k$ and $\mathcal{F}_k$ to the versions in which the gates whose supports are disjoint from the support of the Heisenberg-evolved operator. We shall explain this in more detail in Section~\ref{sec:pcc_transfer_matrix}.

\begin{figure*}[t]
  \subfloat[][$D=2$]
           {
             \includegraphics[width=0.175\textwidth]{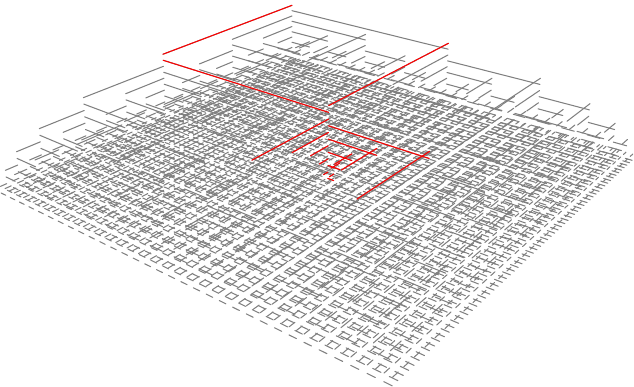}
           }
    \subfloat[][$D=3$]
           {
             \includegraphics[width=0.175\textwidth]{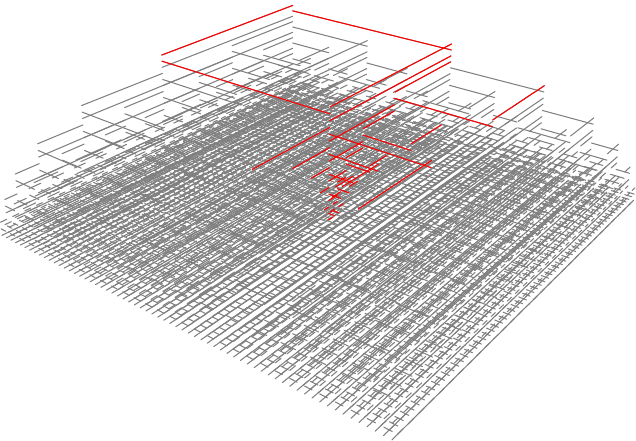}
           }
             \subfloat[][$D=4$]
           {
             \includegraphics[width=0.175\textwidth]{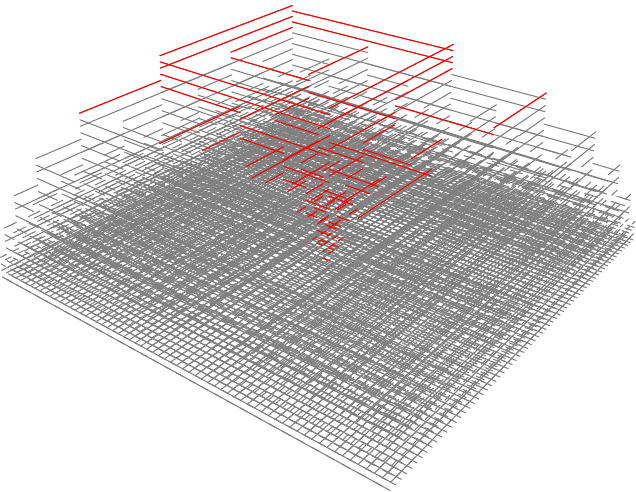}
           }
             \subfloat[][$D=5$]
           {
             \includegraphics[width=0.175\textwidth]{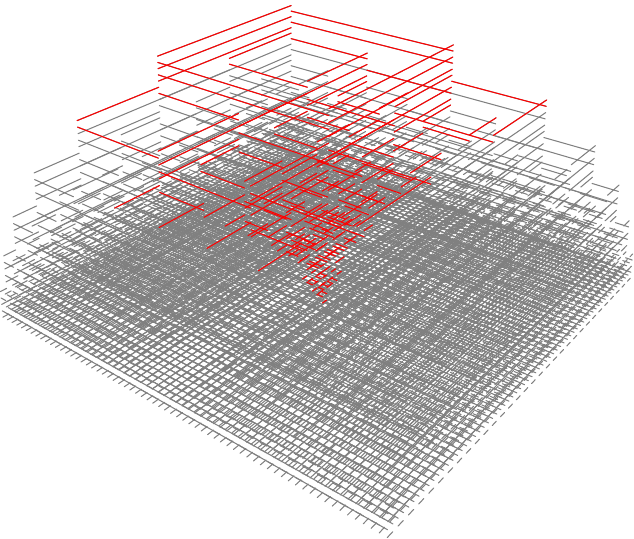}
           }
           \subfloat[][$D=6$]
           {
             \includegraphics[width=0.175\textwidth]{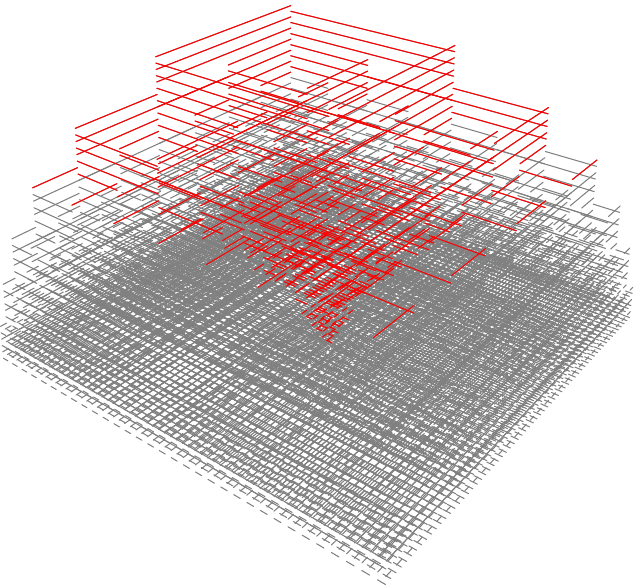}
           }
           \caption{Examples of past causal cone for 2D DMERA. The black lines represent two-qubit gates appearing in the DMERA and the red lines represent the past causal cone of an observable at site $(x,y) = (16, 16)$. Here $n$ is chosen to be $6$.\label{pcc:2d_ex}}
\end{figure*}

\subsection{Transfer matrix}
\label{sec:pcc_transfer_matrix}
The past causal cone discussed in Section~\ref{sec:pcc} can be more conveniently described in terms of \emph{transfer matrices}. Recall that, in constructing the past causal cone, we remove the gates whose supports are disjoint from the support of the Heisenberg-evolved operator. Given an operator $\hat{O}$, this entails changing the $\mathcal{F}_i$ and $\mathcal{U}_i$ to the ones in which these gates are removed, which we shall denote as $\mathcal{F}_i'$ and $\mathcal{U}_i'$ below. Then the expectation value of $\hat{O}$ can be written as
\begin{equation}
\Phi_n \circ  \ldots \circ \Phi_{1} (\hat{O}),
\end{equation}
where $\Phi_k$ can be obtained from the composition of $\mathcal{U}_{n-k}'$ and $\mathcal{F}_{n-k}'$, specifically
\begin{equation}
  \Phi_k(\cdot) = {\mathcal{F}_{n-k}'}^{\dagger} {\mathcal{U}_{n-k}'}^{\dagger}(\cdot) \mathcal{U}_{n-k}' \mathcal{F}_{n-k}'.
\end{equation}
We refer to $\{\Phi_k \}$ as the set of transfer matrices.

In principle, all the transfer matrices can be different, in particular, if all the gates in the DMERA are chosen differently. However, often times it is helpful to choose the gates in a translationally invariant way. Such a choice is motivated from existing numerical studies and known analytic constructions of MERA~\cite{Evenbly2016,Haegeman2017}. If that is the case, a more regular pattern arises within the transfer matrices.

\begin{figure*}[t]
  \includegraphics[width=1.0\textwidth]{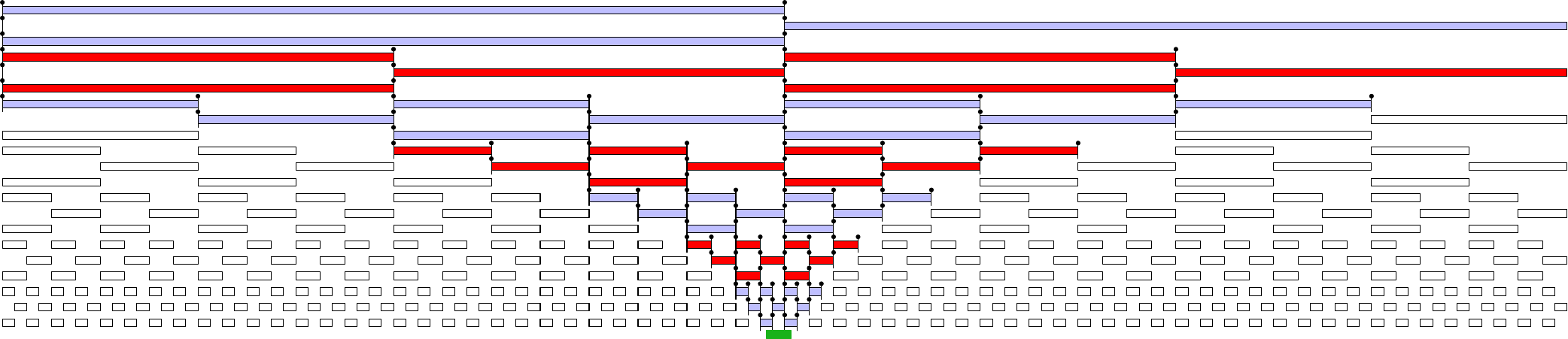}
  
  \includegraphics[width=1.0\textwidth]{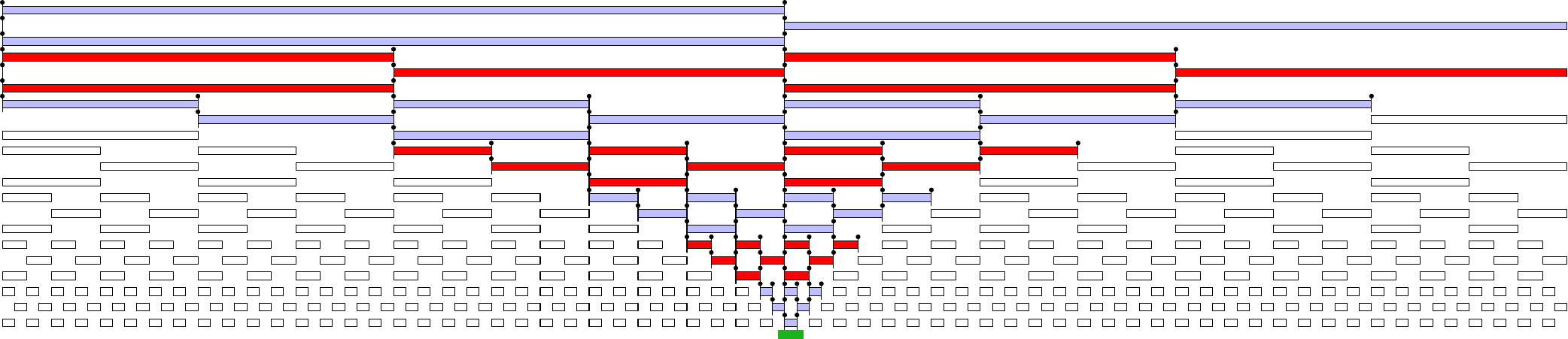}
  
  \includegraphics[width=1.0\textwidth]{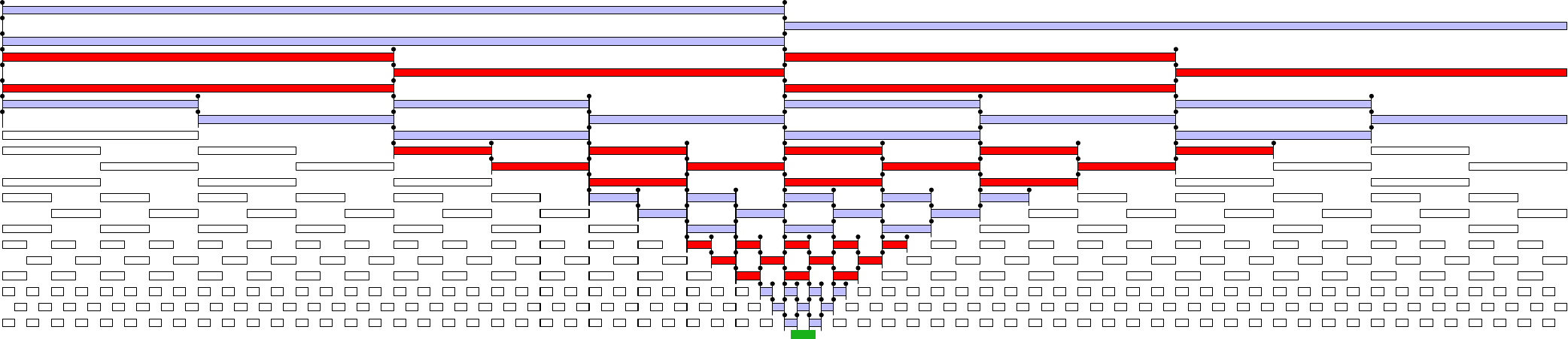}
  
  \includegraphics[width=1.0\textwidth]{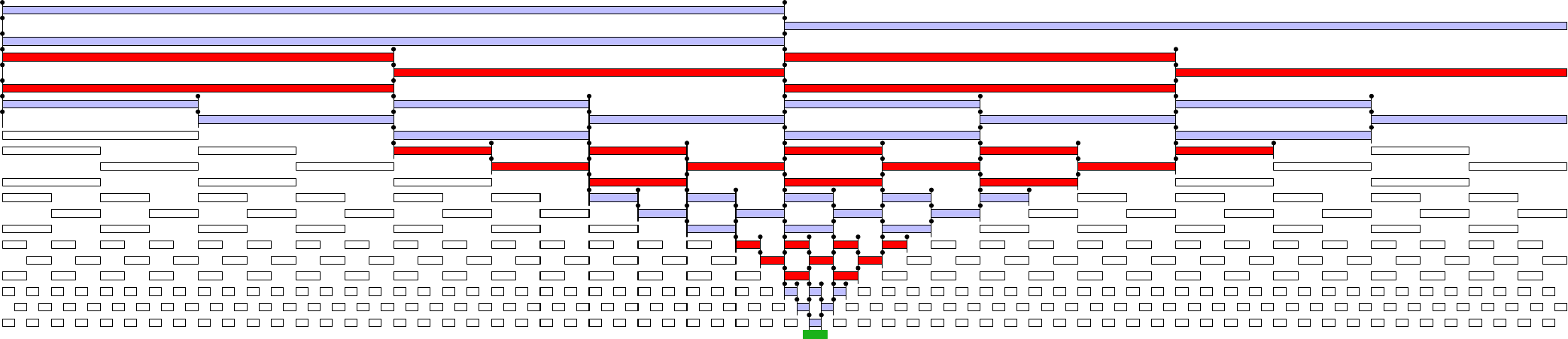}
  
  \caption{Examples of past causal cone for nearest-neighbor observables on $(n=8, D=3)$-DMERA. The nearest-neighbor observable is represented as a green rectangle. Different scales are colored in red/blue depending on whether the scale is odd/even; this is to clearly delineate the different scales. One can see that the transfer matrix within each scale can be obtained by shifting a sufficiently large transfer matrix within that scale and then further restricting that transfer matrix to a smaller set of qubits.\label{fig:pcc_regularity}}
\end{figure*}

To explain how this pattern works, it will be helpful to provide an example of the one-dimensional DMERA. We have computed the past causal cone of nearest-neighbor observables, which are then shifted by an integer multiple of the unit spacing [Fig.~\ref{fig:pcc_regularity}]. As one can see, all the past causal cones within each scale can be obtained from a shift and a restriction of a  transfer matrix on a sufficiently large interval. The same conclusion applies to 2D DMERA. 

The shape and the size of the past causal cone play a fundamental role in the resource estimate for DMERA. For each scale $s$, we denote the maximum width as $w_s$. The maximum of $w_s$ over $s$ is the number of qubits we need to run DMERA.

\section{Gradient scaling}
\label{sec:barren_plateau}

In this Section, we study the typical value of the gradient when the gates chosen in the DMERA are random. To model whether the ansatz generates a barren plateau, namely a gradient that scales polynomially in some variable only for an exponentially small volume of the space of all circuits, we describe our circuit as a product of Haar random two-qubit unitaries, 
\[\prod_{i=N}^1 \mathcal{U}_i \ket{\vec{0}}\]
and take 
\[E = \expval{H} = \mel{0}{\prod_{i=1}^N \mathcal{U}_i^\dagger) H (\prod_{i=N}^1 \mathcal{U}_i)}{0}\]
. Each unitary $\mathcal{U}_i$ is composed itself of a set of unitaries $\mathcal{U}_i = \prod_j U_j$ where j runs over the total number of gates at level $i$.
For simplicitly, we will simply compress all indices and represent our circuit as a long list of unitaries $\{U_i\}$ and denote the number of such unitaries as $N$ for the purposes of this section.
We are interested in taking gradients of this energy with respect to the individual unitaries $U_i$. 
We can define a gradient of one of the gates $U_k$ by perturbing it with a Pauli operator $P$, ie replacing $U_k$ with $U_k \exp{-i\epsilon P}$, and computing
\begin{align*}
  % \begin{aligned}
   \lim_{\epsilon \rightarrow 0} \frac1\epsilon  (& \mel{0}{(\prod_{i=1}^{k-1} U_i^\dagger) e^{-i\epsilon P} (\prod_{i=k}^N U_i^\dagger) H (\prod_{i=N}^k U_i) e^{i\epsilon P} (\prod_{i=k-1}^1 U_i)}{0}\\ &\quad - \mel{0}{(\prod_{i=1}^N U_i^\dagger) H (\prod_{i=N}^1 U_i)}{0})
  % \end{aligned}
\end{align*}

For convenience, we define $\ket{0'} = \prod_{i=k-1}^1 \ket{0}$ and $P_j' = \prod_{i=k}^N U_i^\dagger P_j \prod_{i=N}^k U_i$, and take the Hamiltonian to be some sum of Pauli operators $H = \sum_j \alpha_j P_j$. With this convention the above can be rewritten as
\begin{align*}
  \lim_{\epsilon \rightarrow 0} \frac1\epsilon  \mel{0'}{ (e^{-i\epsilon P} \sum_j \alpha_j P_j' e^{i\epsilon P})-\sum_j \alpha_j P_j' }{0'}
\end{align*}
Expanding the sum over $j$ we get
\begin{align*}
  \partial_{P,k}E &= \sum_j \alpha_j \partial_{P,k}\expval{P_j} \\
  \partial_{P,k}\expval{P_j} &= \lim_{\epsilon \rightarrow 0} \frac1\epsilon  \mel{0'}(e^{-i\epsilon P} P_j' e^{i\epsilon P} - P_j'){0'} \\
  &= \lim_{\epsilon \rightarrow 0} \frac1\epsilon  \mel{0'}{P_j' - i\epsilon  (P_j'P-PP_j') + \order{\epsilon ^2} - P_j'}{0'} \\
  &= \text{Im}(\mel{0'}{\comm{P_j'}{P}}{0'})
\end{align*}

Our interest is the distribution of $\partial_{P,k}E$ over the space of circuits generated by Haar random two-qubit unitaries $U_i$. 
We can see immmediately that $\expval{\partial_{P,k}E}_{~\text{Haar}} = 0$, as for every circuit which generates a positive expectation value for any term one can modify it to produce the negated value. 
Our question then is to understand the variance of $\partial_{P,k}E$, which since the expectation value is zero becomes just 
\begin{align*}
  &\text{Var}(\partial_{P,k}E) = \expval{\partial_{P,k}E}_\text{~Haar} = \sum_{i,j} \alpha_i \alpha_j \expval{\partial_{P,k}P_i \partial_{P,k}P_j}_{\text{Haar}} \\
  &\quad = \sum_{i,j} \alpha_i \alpha_j \expval{\text{Im}(\mel{0'}{\comm{P_i'}{P}}{0'}\text{Im}(\mel{0'}{\comm{P_j'}{P}}{0'}}_{\text{Haar}}
\end{align*}

Note that the expectation value from Haar-random circuits is equal to the expectation value from random Clifford circuits with identical connectivity, and thus we can say that 
\begin{align*}
  \text{Var}(&\partial_{P,k}E) = \\ &\sum_{i,j} \alpha_i \alpha_j \expval{\text{Im}(\mel{0'}{\comm{P_i'}{P}}{0'}\text{Im}(\mel{0'}{\comm{P_j'}{P}}{0'}}_\text{Clifford}
\end{align*}

We note some key properties of the terms $\partial_{P,k}\expval{P_j} = \mel{0'}{\comm{P_j'}{P}}{0'}$. 
Thanks to shifting to drawing random Clifford circuits, we note that $P_j'$ is now another Pauli string as opposed to some general operator, as Cliffords map Paulis into Paulis with some phase. 
Specifically, $P_j'$ is some Pauli string that is supported on the past causal cone of $P_j$ at the point in the circuit corresponding to gate $U_k$. 
Second, $\comm{P_j'}{P}$ thus is either $0$ if operator $P$ commutes with $P_j'$, or it is equal to $2P_j'P$.
Thus, the value of $\partial_{P,k}\expval{P_j}$ for any given circuit is $0$ if $P$ and the image of $P_j$ defined over that operator's PCC at gate $U_k$ commute, or it is equal to twice $\expval{P_j''}{0}$ where $P_j''$ is the Pauli operator corresponding to $P_j'P$ projected back through to the beginning of the Clifford circuit.
This evaluates to $0$ if $P_j''$ has any $X$ or $Y$ Paulis, otherwise it evaluates to $1$.

Thus, we can recast the task of computing the variance of the gradient of the energy to simply asking what the probability is that a) $P_j'$ and some $P$ commute and b) that $P_j''$ has no $X$ or $Y$ Pauli operators, and what the correlation is between that probability for different Hamiltonian terms $P_j$.

\subsection{Upper bound of the variance of the gradient}
We can upperbound the probability by first taking the norm of each term in the variance, to discard consideration of the phase coming from the Clifford circuit. 
Second, we can pull out a factor of $4$ and treat each term in the product as a simple binary variable (corresponding in value to whether or not both $\comm{P_j'}{P}\neq 0$ and if $P_j''$ contains no $X$ or $Y$ Paulis). 
Third, we note that $\expval{x_i x_j}$ for some binary variables $x_i$ is strictly smaller than $\max_{i,j}(\expval{x_i},\expval{x_j})$. 
Thus, we can upperbound the variance by asking this question: What is the probability that both our conditions $\comm{P_j'}{P}\neq 0$ and $\{\sigma_X,\sigma_Y\} \notin P_j''$ are satisfied for a random Clifford circuit?
If this probability decays exponentially in either the DMERA depth $D$ or the number of layers $n$ then we are guaranteed to have a barren plateau along that parameter and the circuit will rapidly become untrainable as we scale that parameter.

The last important feature to note is that a random 2-qubit Clifford operation maps $II \rightarrow II$ while transforming any other Pauli string to a random Pauli string (that is not $II$). 
We can then see that the action of the circuit on a Pauli string operator $P$ is the identity on all gates that do not interact with $P$, while scrambling the Pauli string otherwise.

Let us examine the case with two gates $G_1$ and $G_2$ acting on a set of 3 qubits $q_1,q_2,$ and $q_3$, where $q_1$ and $q_2$ are inside our Pauli string and $q_3$ is just outside it. Assume that the Pauli string is not $II$ on $q_1,q_2$, then $G_1$ will scramble the string and the operator on $q_2$ will be $I$ with probability $3/15$. If this is the case, then the gate $G_2$ will be acting on $II$ and not expand the Pauli string.

There are two categories of operators, $I$ and $\{X,Y,Z\}$, and the action of random Cliffords renders $\{X,Y,Z\}$ equally likely at every stage, and we can simplify our discussion by only looking at a binary choice between these two classes, call them $I$ and $V$.
Finally, we note that for our purposes of placing a bound on the variance, we can ignore the question of whether the $\comm{P_j'}{P}=0$ or not as it is at most a constant factor (since $P$ is just a two-body operator), and moreover that since the Clifford circuit randomizes $P_j'$ we can drop dependence on the specific Pauli $P$.

Now, bounding the variance is equivalent to asking what is the distribution of Pauli operators in $P_j''$.

We can model the evolution of $P_j$ through the random Clifford circuit as a random walk. ``In the bulk'' of the past causal cone, individual operators are $I$ with a probability of $3/15$, as each gate acting in the bulk has a probability of $3/15$ of setting each of the operators to $I$ (but not both). Therefore, the probability of all operators in $P_j''$ being $I$ or $Z$ is driven by the size of the boundary of support for $P_j''$.

As demonstrated above, a gate acting at the boundary of the PCC can either expand it (with probability $12/15$) or contract it (with probability $3/15$). Extinction of $P_j$'s PCC is impossible as random Cliffords cannot map any operator to $II$ except $II$. 
We can treat the movement of the boundary as a kind of random walk. 
In a single dimension, take the width of the effective PCC after each round of nearest neighbor interactions in the DMERA as a random walker that can move $-2,-1,0,1,$ or $2$ steps on each move, bounded from below at $2$ (as $P_j$ is a two-body operator) with probabilities of $1/25,2/25,6/25,6/25,$ and $9/25$ respectively.
To see this, simply count the possibilities and probability of each outcome from the fact that each edge has a $3/15$ chance of placing an $I$ on the boundary and thus pulling the PCC inward, a $3/15$ chance of placing some other operator on the boundary and thus not moving it, or otherwise expanding the PCC.

This random walk will grow with time at an average rate of $4/5$ per round, with a standard deviation which grows as the square-root of the number of rounds. After $D$ such rounds, we thus expect the PCC's operator along one dimension to extend to $0.8D$ qubits. After each layer of our DMERA circuit is completed, we then reduce this width by a factor $r$, and continue to expand.

In expectation, then, the width in a single dimension of $P_j''$ after pushing $P_j$ backward through an entire random Clifford circuit will be, as one could predict from Eq. \ref{eq:pcc_width}, $\expval{\text{Width}} = 0.8 Dr/(r-1)$ with a standard deviation that scales as $\sigma = \order{\sqrt{D}}$. 

Thus, the ball corresponding to the PCC of $P_j$ has a radius of $R \approx 0.4 Dr/(r-1)$, and its volume scales as $R^d$. After each layer of DMERA, an additional volume \[\order{(R(1-1/r))^d}=\order{(0.4D)^d}\] with $n$ such levels for a total number of qubits $\order{(0.4D)^d n}$ inside $P_j''$ in expectation.

However, while on average the volume of the PCC grows linearly in $n$ and thus a large fraction of all circuits will contribute $\order{exp(-n)}$ to $\expval{P_j}$, at each scale level, label it $s\in [1,n]$ there is the possibility that $P_j'$ will have support only on the $\order{(0.4D)^d}$ nodes which will not be acted upon again later in the circuit -- that is, the Pauli operator $P_j'$ will freeze and no longer evolve as we apply the rest of the gates in the PCC of $P_j$. The probability that this occurs does not scale with $n$, but rather only with the volume of the PCC after one layer, $\order{D^d}$. Therefore, there is a constant fraction of all Clifford circuits as a function of $n$ will have a $P_j''$ with support on these $\order{D^d}$ nodes.

Moreover, while we expect that the vast majority of circuits will have support on $\order{D^d n}$ nodes, there is a small probability that $P_j'$ does not percolate through the graph at all -- at each layer it has a some chance of not expanding, and thus it has a probability $p = \order{exp(-Dds)}$ of simply not expanding through scale $s$, and remaining in a constant size region of support approximately equal to that of $P_j$. These terms individually make the largest contributions, as they have a large ($\approx$ constant) probability of containing no X or Y Paulis and that probability decays as $3^{-K}$ for a Pauli with $K$ non-identity operators (the more general case for the weight of the Pauli string up to scale $s$ would have a probability of $3^{-\order{D^d s}}\ll exp(-\order{Dds})$ for $d>1$). The expectation value for $\mel{0}{P_j''}{0}$ is thus dominated by these components. The contributions from these paths can be upper-bounded by assuming the resulting Pauli strings evolve, as in the previous paragraph, to only have support on qubits which are not acted on at longer length scales. However since these $P_j'$'s are small of constant size, this probability is $p^{-K}$ with $p$ on the order of $\frac{r^d-1}{r^d}$ for a DMERA circuit that grows by a factor of $r$ at each scale.

Taking the derivative, then, of a parameter present in scale $s$, we find the derivative is actually dominated by the cases where $P_j'$ has not expanded in support significantly beyond the size of $P_j$, and where it does not expand significantly as it evolves to $P_j''$, ie terms whose contributions decay exponentially with $Dds$ (the number of gates each qubit participates in in a layer) rather than the naive $D^d n$. 

Each of these non-$I$ operators is a random Pauli equally distributed over $\{X,Y,Z\}$, then 
\begin{equation}
  \text{Var} \partial_{k,P} \expval{P_j} \leq \order{\exp(-Dds)}
\end{equation}
and therefore by Chebyshev's inequality the volume of all circuits with polynomially scaling variance for DMERA is exponentially small in the depth $D$.

Thus, we find that our upper bound allows gradients to scale polynomially with system size, but they decay exponentially with depth parameter $D$, indicating a barren plateau in $D$.

\subsection{Lower bound}

We can go further than our upper bound by investigating the behavior of $\expval{\text{Im}(\mel{0'}{\comm{P_i'}{P}}{0'}\text{Im}(\mel{0'}{\comm{P_j'}{P}}{0'}}_{\mathstrut \text{Clifford}}$. 

First, we will note that a random Clifford gate acting on a Pauli string will scramble the phase coefficient in front of the Pauli string so long as it does not act trivially (ie as long as the Pauli is not the identity on the support of the gate). The specific properties of the cross-correlation terms, $i \neq j$, will depend on the parameters $\alpha$. Moreover, for any given Clifford circuit, one can replace the next gate to act on the support of $P_i'P$ so as to have the opposite phase, resulting in a cancellation of these terms as we average over the ensemble of all Clifford circuits. However, the terms of the form $\alpha_i^2 \expval{\text{Im}(\mel{0'}{\comm{P_i'}{P}}{0'})^2}_{\text{Clifford}}$are independent of the sign of the phase, and so as above in the upper-bound case reduces to simply taking an expectation value over an ensemble of terms that are either 0 or 1 over all Clifford circuits, with no cancellation possible as in the $i \neq j$ case.

As in the upper bound case, the scrambling properties of Cliffords acting on Pauli strings allows us to drop specific dependence on $P$, and we again see the uniformly random transformation over all gates acting non-trivially on a Pauli string allows us to simplify our construction into two classes of operators: at any given location the $P_i'$ has either the identity at any given site or a uniformly random selection over $\{X,Y,Z\}$. One can see immediately that the arguments above, mapping the expansion of the evolution of the support of the $P_i'$ operator through the PCC of $P_i$ acts a random walk, go through unchanged. 

We can conclude that for the purposes of a lower bound, we can ignore the contributions of large-support $P_i''$ circuits (since each circuit can only contribute a positive value to the expectation) and observe that we have a probability that scales as $\order{\exp(-c D d s)} = \order{\exp{-cD d\log{N_s}}}=\order{\exp{-cD}N_s^d}$ for a positive value $0<c<1$ for the support of the PCC to not expand beyond the region of $P_i$ up to scale $s$ and volume up to scale $s$ labeled $N_s$.

We see that the upper bound and lower bound both have the same form with respect to $D$ and $s$, that is, depth and the scale at the location of the target parameter (at most log of the system linear size).

We note that this result may appear to differ with \cite{Barthel2023absence}, however we note that that reference is concerned with MERA circuits, rather than DMERA. As a result, at each scale, there is only a constant number of operations with neighbors, effectively $D=1$, and our result here agrees that there should be no barren plateau in that case in system size $N$. DMERA is a more general case, and we observe an exponential suppression of the variance of the gradients with depth parameter $D$.

\subsection{Numerical simulations}
To demonstrate this, we simulate the evolution of a Pauli operator backward through the DMERA circuit in 2D in a simplified setting, modeling only the effect of random Clifford gates according to the Monte Carlo update rule above, namely that a random Clifford gate will map $II \rightarrow II$ with $p=1$ and fully scramble all other Pauli operators with equal probability. Thus we need only model whether the operator is $I$ or a random Pauli $V\in\{X,Y,Z\}$. We thus can select a pair of qubits denoting a measurement at the end of a DMERA circuit, assign the support of the measurement the state of being some non-$I$ Pauli $V$ (denoting a random $X,Y$ or $Z$ Pauli), and then evolve that state back through a random Clifford circuit to find the effective template for a realization of the Pauli PCC of the measurement, denoted above by $P_j''$. 

Describing these templates for $P_j''$ in a binary string $\vec{K_{P_j}}$, with $0$ denoting an identity gate and $1$ denoting a random Pauli, we can then simply evaluate the Hamming weight of the binary string, $\abs{\vec{K_{P_j}}}$, and look at the distribution of $ \abs{\vec{K_{P_j}}}$ as a function of $n$ and $D$. We suppress the dependence on $P_j$ for the time being for easy of notation. 

A random instantiation of a template Pauli string $\vec{K}$, is found by building a Pauli string with $I$ everywhere $K_i=0$ and a random value $X,Y,$ or $Z$ everywhere $K_i=1$. The probability that $\mel{0}{\vec{K}}{0}=1$ (it can only equal 1 or 0) is simply $3^{-\abs{\vec{K}}}$. Thus, given an ensemble of random such templates one can estimate the value of $\mathbb{E}[\mel{0}{P_j''}{0}]_{\textrm{Clifford}}$, our upper bound on the gradient, as 

\begin{equation}
  \label{eq:upperboundexpval}
  \mathbb{E}[\mel{0}{P_j''}{0}]_{~\textrm{Clifford}} = \sum_k Pr(\abs{\vec{K_{P_j}}}=k) 3^{-k}
\end{equation}

By sampling many templates for a given measurement operator's support we can estimate our upper-bound the $\partial_{P,k} \expval{P_j}$. We can also give an average value of our upper bound over the measurements for each operator.

We simulate this process to construct $10^6$ templates for $P_j''$ for each of 100 randomly selected (with replacement) measurement operators (defined by their supports) for each combination of $n\in\{3,4,5,6,7,8\}$ and $D\in\{1,2,3,4,5,6\}$ in two dimensions and present the results below.

\begin{figure}
  \centering
  \includegraphics[width=\columnwidth]{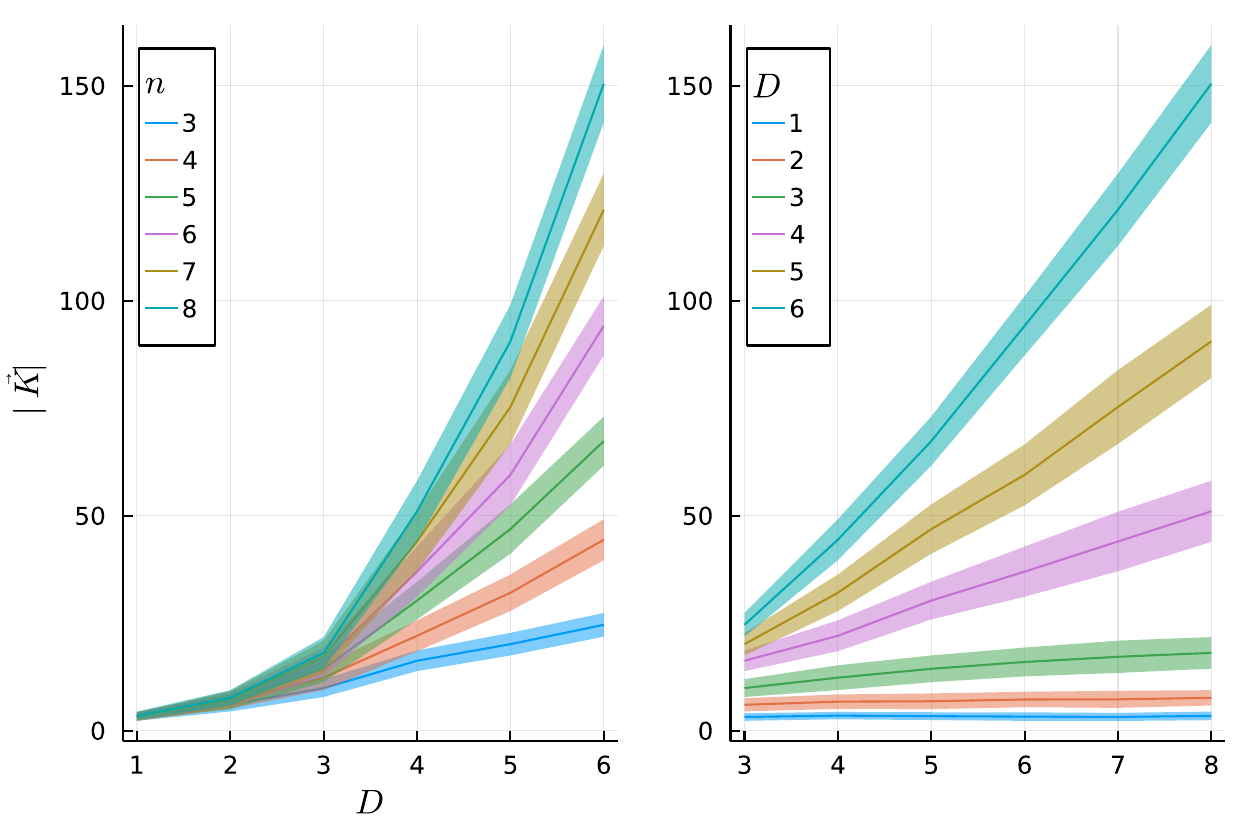}
  \caption{We plot the expectation value and standard deviation of $\abs{\vec{K}}$ as a function of $D$ and $n$, and we see our effective past causal cones on average scale in volume polynomially with $D$ but linearly with $n$, as expected.}
  \label{fig:absvecK_vs_D_n}
\end{figure}

First, we simply will plot $\abs{\vec{K}}$ as a function of $D$ and $n$ in Fig. \ref{fig:absvecK_vs_D_n}, to verify our expectation that on average it scales quadratically and linearly, and it does as expected.

\begin{figure}
  \centering
  \includegraphics[width=\columnwidth]{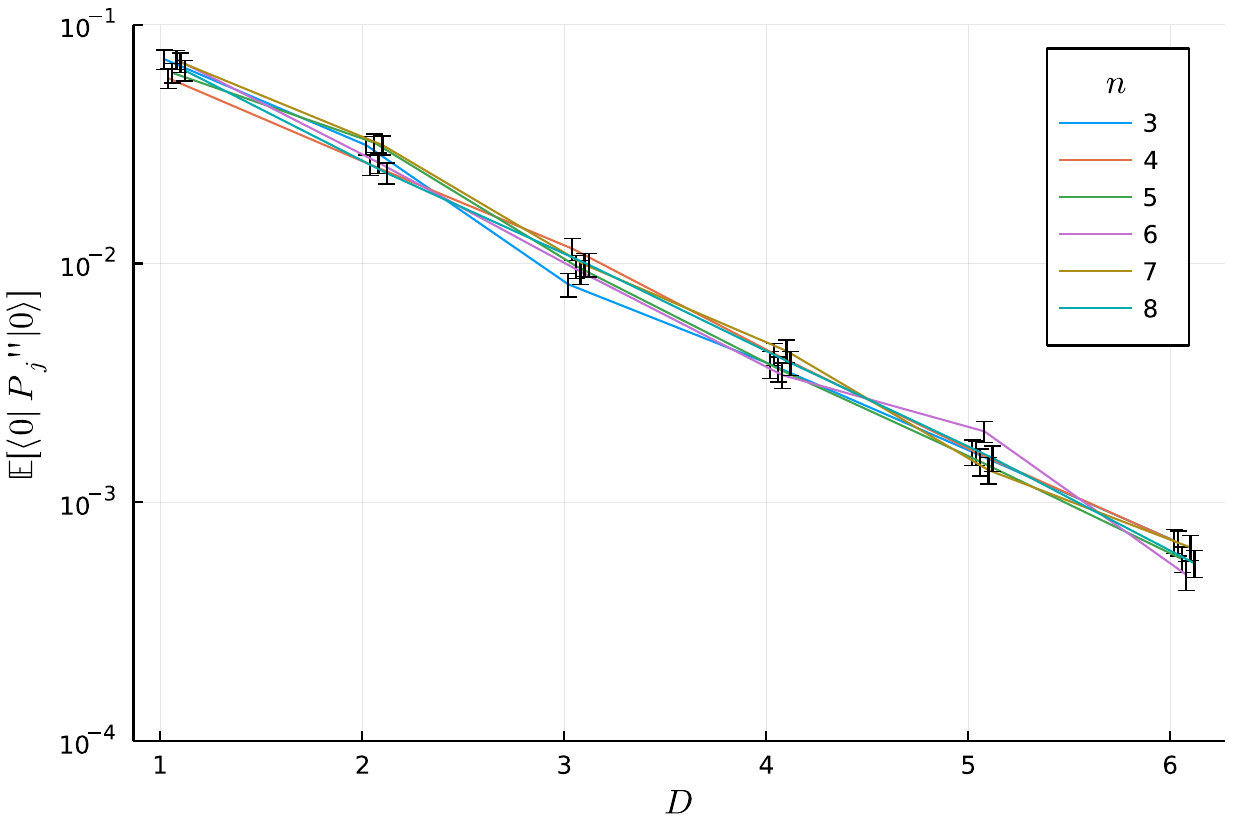}
  \caption{The expectation value of our upper bound on $\partial_{P,k}\expval{P_j}$ over $100$ randomly selected $P_j$, ribbon denotes the $1\sigma$ of the expectation values, resulting from a Bayesian boostrap. We see very clearly that there is no relationship between this and $n$, as predicted, while there is a strong exponential dependence on $D$.}
  \label{fig:expval_ribbon_D_by_n}
\end{figure}

Next, we take the expectation value of our upper bound, as defined in Eq. \ref{eq:upperboundexpval} over all selected measurement operators along with one sigma error bars for said expectation value (determined via a Bayesian bootstrap \cite{rubin1981bayesian} over the sampled measurement operators) and plot it as a function of $n$ and $D$ in \ref{fig:expval_ribbon_D_by_n}, showing a clear exponential decline with $D$ and no $n$ dependence, as theory predicts due to the heavy suppression of high-weight random Pauli strings due to low probability of such strings containing no $X$ or $Y$ Pauli operators. This leaves the region in which our Pauli string does not expand significantly over time over the circuit as the dominant contributor to our upper bound.

\begin{figure}
  \centering
  \includegraphics[width=\columnwidth]{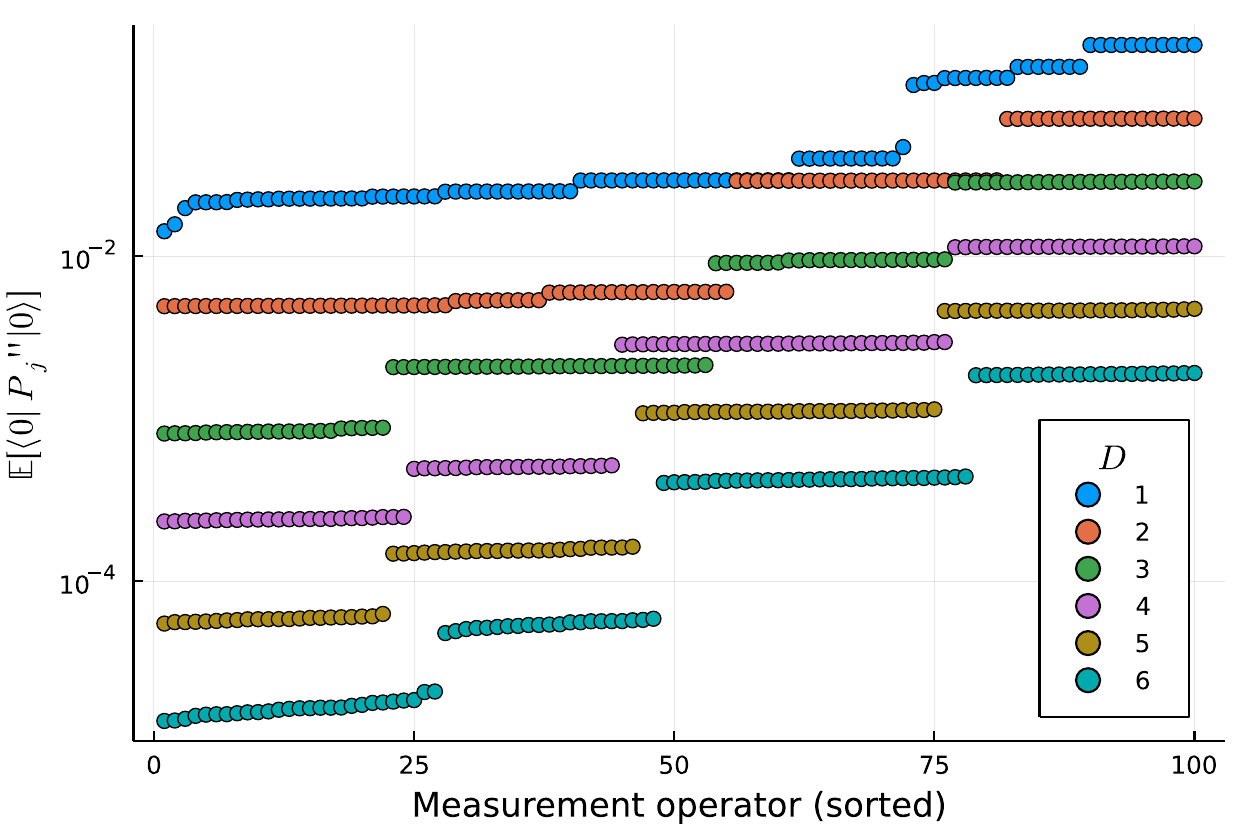}
  \caption{The expectation value of our upper bound on $\partial_{P,k}\expval{P_j}$ over $100$ randomly selected $P_j$ for various $D$ and $n=8$, sorted along the x-axis from least to greatest. We see discretized behavior, which we attribute to varying probability to observe the very smallest $\abs{\vec{K}}$}
  \label{fig:scatterupperbound}
\end{figure}

We may also break out the individual values for our upper bound for each of the 100 measurement operators for different depths $D$, for a fixed $n=8$, though as is apparent from Fig \ref{fig:expval_ribbon_D_by_n}, there is no significant dependence on $n$. We show a scatter plot of all such values, sorted by value of the upper bound, in Fig \ref{fig:scatterupperbound}. The discrete bands we see with limited variation are attributable to different probabilities of observing the smallest possible $\abs{\vec{K}}$.

We observe clearly that while there is a large separation across measurement operators, as a function of D there is a consistent exponential decline in the upper bound on $\partial_{P,k}\expval{P_j}$. Thus, as predicted, there exists a barren plateau for $D$, but negligible dependence of the upper bound for the gradient on $n$.

\section{Resource estimate: Incoherent approach}
\label{sec:resource_estimate_incoherent}

In this Section, we quantify the resource required to run a DMERA circuit in terms of the number of qubits, circuit depth, and the number of one- and two-qubit gates. We assume the standard model of quantum computation in which two-qubit gate between any pair of qubits is possible, and we assume the ability to perform reset of the qubits. 

\begin{figure}[h]
  \includegraphics[width=0.95\columnwidth]{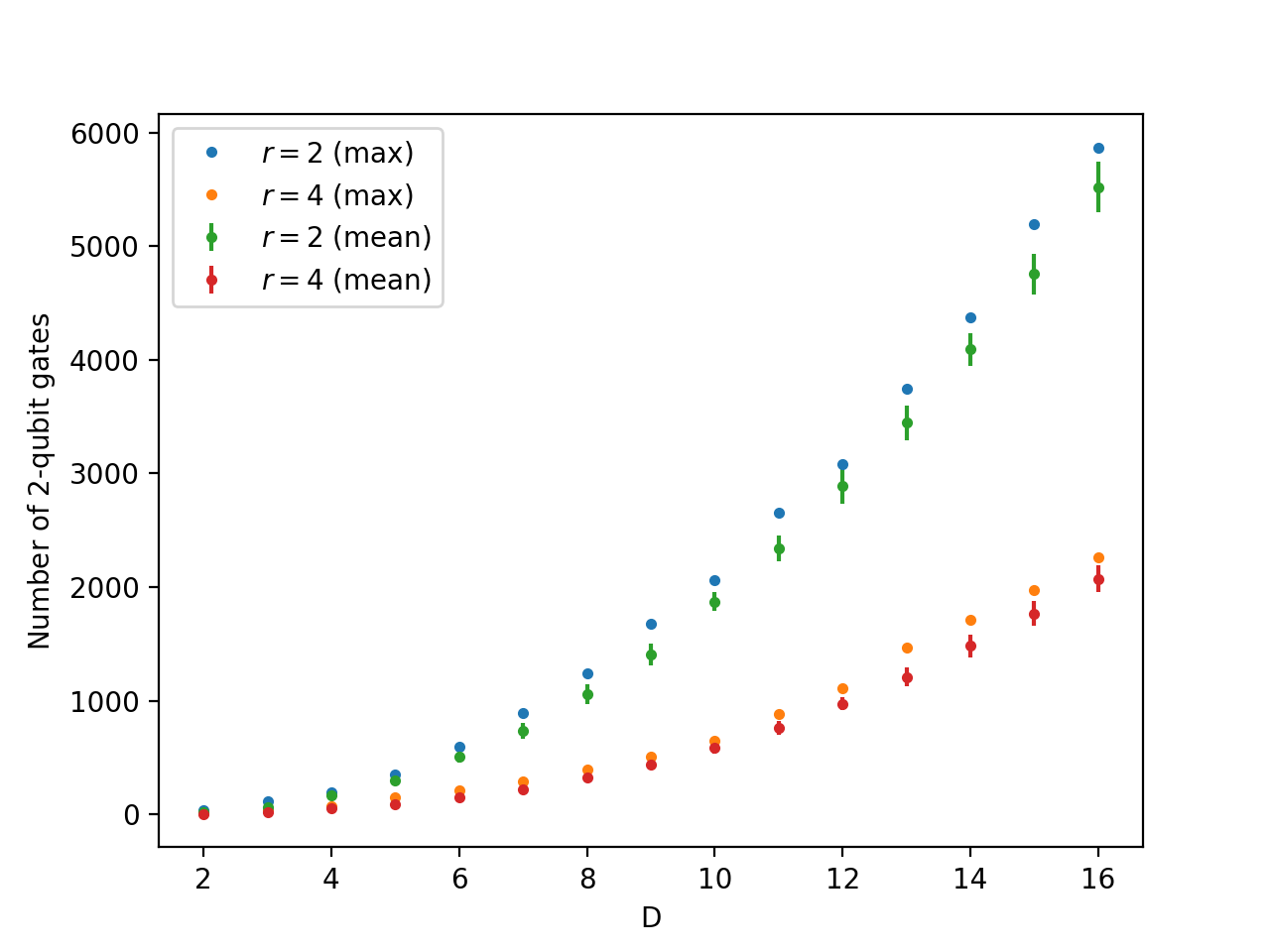}
  \caption{The number of $2$-qubit gates used in DMERA in terms of $D$.\label{fig:D_vs_vol}}
\end{figure}

Let us first note that the circuit depth of the $(n,D)$-DMERA is exactly $nD$ and the number of qubits needed is upper bounded by $\mathcal{O}(D^d)$ and the number of gate is $\mathcal{O}(nD^{d+1})$, where $d$ is the number of spatial dimensions. However, in resource estimation the precise constants matter. We perform a detailed analysis by explicitly computing the past causal cone and estimating its width. Depending on how this circuit is compiled, the depth and the width of the circuit can change. However, the total number of gates remains invariant. We have plotted this number in terms of $D$ for $D$ ranging from $2$ to $16$ in Fig.~\ref{fig:D_vs_vol}. For concreteness, we chose the system size to be $64\times 64$, motivated from the physical examples discussed in the introduction.

\begin{figure}[h]
  \includegraphics[width=0.95\columnwidth]{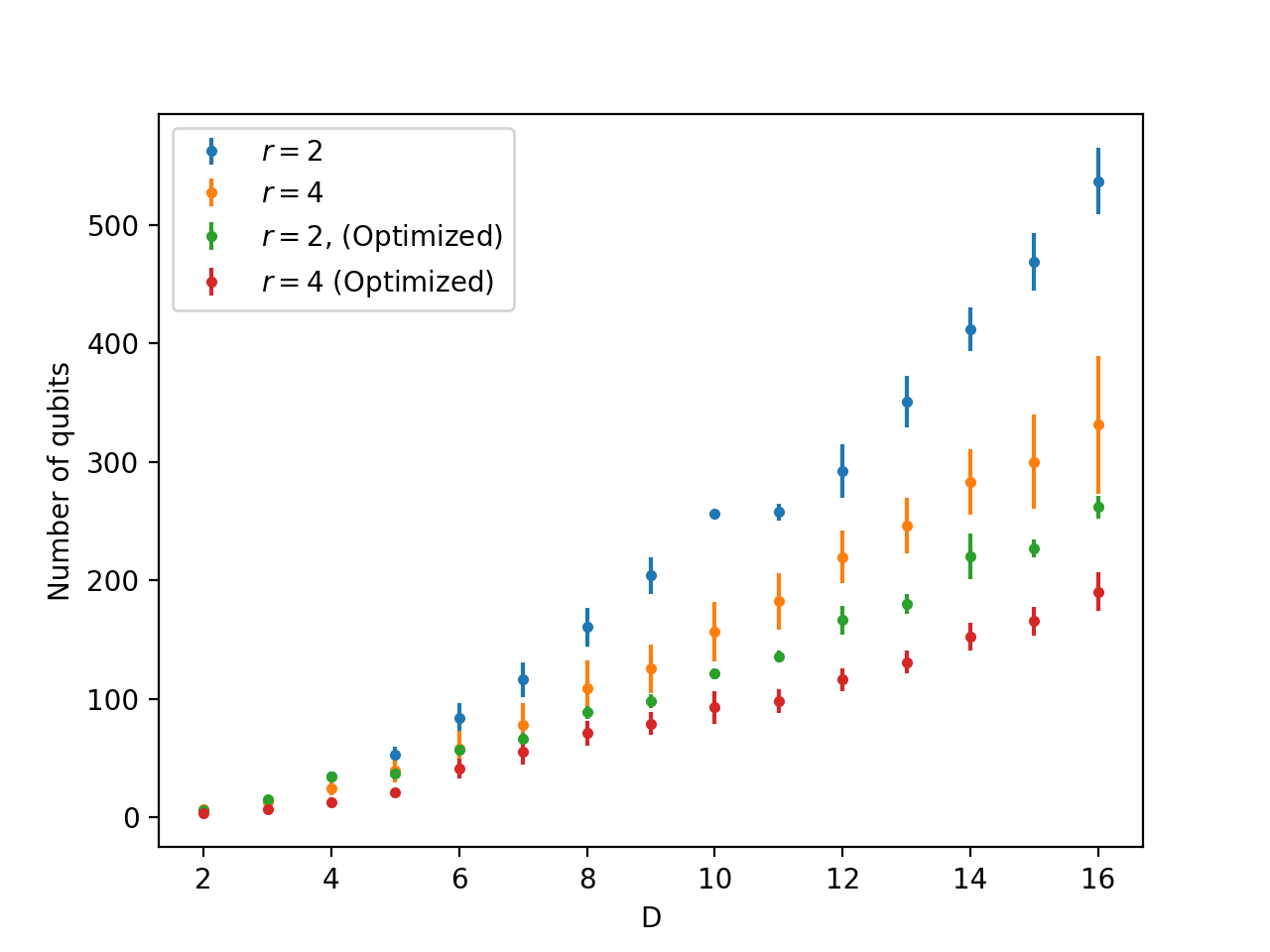}
  \caption{The number of qubits needed for computing past causal cone. Each data point was obtained from $100$ randomly generated nearest-neighbor local observables on a two-dimensional lattice. Generally, larger value of $r$ (see Section~\ref{sec:pcc}) reduces the qubit cost. The compilation technique in Appendix~\ref{appendix:space_efficient_compilation} also led to a noticable reduction in qubit cost.\label{fig:dmera_incoherent_cost}}
\end{figure}

We have also employed a novel circuit optimization technique that can reduce the number of qubits substantially. We introduce a novel compilation technique which we describe in detail in Appendix~\ref{appendix:space_efficient_compilation}. This technique reduces the overall number of qubits needed with the expense of increasing the depth of the circuit. The comparison between these two are made in Fig.~\ref{fig:dmera_incoherent_cost}.

In a fault-tolerant quantum computer, the relevent figure of merit is the number of non-Clifford gates, such as the $T$-gate. The precise number can be estimated conveniently in terms of two parameters: (i) the number of two-qubit gates in the past causal cone and (ii) the amortized $T$-gate cost for each gate. While (i) depends only on $D$, as shown already in Fig.~\ref{fig:D_vs_vol}, the second number depends on the choice of the gates. On one hand, if one were to use arbitrary two-qubit gates that define the DMERA ansatz, one would need to employ a decomposition of a general unitary in $SU(4)$ into CNOT gates and single-qubit rotations~\cite{Vidal2004}. This entails using at most $4$ arbitrary single-qubit rotations plus $3$ single-axis rotations, accompanied with Clifford gates. Since arbitrary single-qubit rotation can be decomposed further into $3$ single-axis rotation, using at most $15$ single-axis single-qubit rotations suffice. Each such rotation ought to be decomposed further into Clifford and $T$-gates, with the $T$-gate cost ranging anywhere between $10$ to $50$ using standard methods~\cite{Bocharov2015,Kliuchnikov2013,Gosset2014,Bocharov2015a,Ross2016}. Thus, a pessimistic estimate on this number can be as large as $750$ $T$-gates per individual two-qubit gate.

However, in practice, it is highly unlikely that one would employ such structureless gates. Recent studies showed that simple choice of gates can be often used in practice~\cite{Olund2020,Anand2022}. In particular, a two-qubit gate chosen from a trotterized form of the local terms of the Hamiltonian is particularly promising~\cite{Miao2021,Miao2023a}. Moreover, often the gates used in these systems are completely arbitrary but have some uniform structure, such as translational invariance~\cite{Evenbly2016,Haegeman2017}. Under those circumstances, it is possible to apply identical rotations over $n$ qubits with the $T$-gate cost of $4n+\mathcal{O}(\log n  \log (1/\epsilon))$, where $\epsilon$ is the target precision~\cite{Gidney2018}, which in the large $n$ limit, approaches $4$ $T$-gates per rotation.

To summarize, a single coherent run of the DMERA in two spatial dimensions is $n_{2Q}C_{T}$, where $n_{2Q}$ is the number of two-qubit gates, scaling cubically with $D$ [Fig.~\ref{fig:D_vs_vol}] and $C_T$ is the amortized $T$-gate cost of the individual two-qubit gates, ranging between $4$ and approximately $750$.

\subsection{Sample complexity}
\label{sec:sample_complexity}

In this Section, we estimate the sample complexity of estimating a variational energy density up to an additive precision $\epsilon$. Without loss of generality, consider the Hamiltonian of the following form.
\begin{equation}
    H = \sum_{x,y=1}^{L} \sum_k \left(\alpha_k O_{(x,y), (x+1, y)}^{(k)} +\beta_k O_{(x,y), (x,y+1)}^{(k)} \right),\label{eq:hamiltonian_general_form}
\end{equation}
where $\alpha_k, \beta_k\in \mathbb{R}$ are numerical coefficients and $O_{(x,y), (x+1, y)}^{(k)}$ and $O_{(x,y), (x+1, y)}^{(k)}$ are hermitian operators of norm at most $1$. Note that Eq.~\eqref{eq:hamiltonian_general_form} is the most general form of translationally invariant Hamiltonian consisting of nearest-neighbor terms. 

Using translational invariance, the total energy \emph{per site} can be written as:
\begin{equation}
\begin{aligned}
    \frac{\langle H\rangle}{L^2} &= \sum_k \mathbb{E}_{x,y}[ \alpha_k \langle O_{(x,y), (x+1, y)}^{(k)} \rangle
    +\beta_k \langle O_{(x,y), (x,y+1)}^{(k)} \rangle].
\end{aligned}
\label{eq:sampling_over_sites}
\end{equation}
Assuming the individual observables $O_{(x,y), (x+1, y)}^{(k)}$ and $O_{(x,y), (x, y+1)}^{(k)}$ are measured, upon optimizing the number of measurements, we obtain the total number of measurements equal to~\cite{Wecker2015}:
\begin{equation}
  M = \frac{\left(\sum_k (|\alpha_k| + |\beta_k|)\right)^2}{\epsilon^2}.\label{eq:sample_complexity}
\end{equation}

\subsection{Case study: 2D Fermi-Hubbard Model}
\label{sec:fh_model_incoherent}

Now we apply the resource estimate obtained in Section~\ref{sec:resource_estimate_incoherent} and~\ref{sec:sample_complexity} to a concrete physical model of interest: the Fermi-Hubbard Model on a two-dimensional square lattice. Our goal is to compare the resource estimate for our approach against other approaches based on QPE applied to the same model~\cite{Kivlichan2020,Campbell2021}. More precisely, our goal in this Section is to estimate the number $T$-gates necessary to estimate the variational energy of $H'$, up to an accuracy of half a per cent of the total system energy~\cite{LeBlanc2015,Kivlichan2020,Campbell2021} (corresponding to roughly $0.0051L^2$).

The Hamiltonian of the Hubbard model is defined as
\begin{equation}
  H = t\sum_{\sigma \in \{\uparrow, \downarrow \}}\sum_{\langle i, j\rangle} a_{i,\sigma}^{\dagger}a_{j,\sigma} + h.c. + u\sum_i n_{i,\uparrow} n_{i, \downarrow},
  \label{eq:fermi_hubbard}
\end{equation}
where $a_{i,\sigma}^{\dagger}$ and $a_{i,\sigma}$ are fermion creation/annihilation operators and $n_{i,\sigma} = a_{i,\sigma}^{\dagger}a_{i,\sigma}$. Here the summation over $\langle i, j\rangle$ means the summation over $i$ and $j$ which are nearest neighbors on the square lattice.

As pointed out in Ref.~\cite{Campbell2021}, Eq.~\eqref{eq:fermi_hubbard} can be rewritten in the following form:
\begin{equation}
  H = H' - \frac{u}{2}N + \frac{u}{4}I,\label{eq:fermi_hubbard_modified}
\end{equation}
where
\begin{equation}
  H' = t\sum_{\sigma \in \{\uparrow, \downarrow \}}\sum_{\langle i, j\rangle} (a_{i,\sigma}^{\dagger}a_{j,\sigma} + h. c.) + \frac{u}{4}\sum_i z_{i,\uparrow} z_{i, \downarrow}.
\end{equation}
Here $I$ is the identity operator, $N:=\sum_i n_{\uparrow, i} + n_{\downarrow, i}$ is the total electron number, and $z_{i,\sigma} := 2n_{i,\sigma} - I$. The last two terms in Eq.~\eqref{eq:fermi_hubbard_modified} commute with $H'$, and as such, they can be simultaneously diagonalized. Moreover, if the underlying state $|\psi\rangle$ is an eigenstate of $N$, e.g., $N|\psi\rangle = \eta |\psi\rangle$, the eigenvalue of $H$ can be obtained straightforwardly from the eigenvalue of $H'$ by subtracting $u(\eta / 2 - 1/4)$. Therefore, for the purpose of computing the ground state energy of $H$ (given a definite value of $\eta$), it suffices to compute the ground state energy of $H'$. 

Upon applying the Jordan-Wigner transformation, the first term in $H'$, for each $\sigma$ and $(i,j)$, can be mapped to an operator of norm at most $t$. Therefore, the estimate for the sample complexity [Eq.~\eqref{eq:sample_complexity}] becomes
\begin{equation}
  \begin{aligned}
    M &\leq \frac{\left(2t+ \frac{u}{4}\right)^2}{\epsilon^2}\\
    &\approx \begin{cases}
      3.5 \times 10^{5} \quad \text{for } t=1, u=4, \\
      6.2\times 10^{5} \quad \text{for }t=1, u=8.
     \end{cases}
  \end{aligned}
  \label{eq:sample_complexity_fh}
\end{equation}
Therefore, for achieving the accuracy of a half a per cent of the total sytem energy, the number of samples needed ranges betweem $3.5\times 10^5$ and $6.2\times 10^5$.

As reviewed in Section~\ref{sec:resource_estimate_incoherent}, in order to estimate $T$-gate cost, one must multiply the numbers in Eq.~\eqref{eq:sample_complexity_fh} by $n_{2Q} C_T$, where $n_{2Q}$ is the number of two-qubit gates being used and $C_T$ is the amortized cost of the individual two-qubit gates. Even under an optimistic assumption that $C_T$ is very close to the lower bound of $4$, the number of $T$-gates obtained this way easily exceeds the number of $T$-gates used in the state-of-the-art methods for QPE~\cite{Campbell2021}, which approaches at most $\approx 2\times 10^6$. Therefore, the incoherent approach is likely to be strictly inferior compared to the QPE-based approaches, in terms of the number of $T$-gates.

One potential advantage of the incoherent approach, however, is the small number of qubits needed. For instance, in Fig.~\ref{fig:dmera_incoherent_cost}, the number of qubits needed remains at least an order of magnitude smaller than the system size ($64\times 64$). Therefore, there may be a merit in using the incoherent approach for DMERA in a small fault-tolerant quantum computer. However, if a large enough quantum computer is available, QPE should be the preferred method of choice.

\section{Amplitude estimation for DMERA}
\label{sec:amplitude_estimation}

A fundamental limitation of the approach discussed in Section~\ref{sec:resource_estimate_incoherent} lies in its scaling in the precision. The sheer number of samples needed to attain a desired accuracy was comparable to the number of $T$-gates needed for the state-of-the-art approaches based on QPE~\cite{Kivlichan2020,Campbell2021}. The approach taken in Section~\ref{sec:resource_estimate_incoherent} is based on samplings of observables, whose scaling in terms of the statistical error $\epsilon$ is $\Omega(1/\epsilon^2)$.

It is a well-known fact that, using the quantum amplitude estimation algorithm~\cite{Brassard2002}, one can improve the scaling to $\mathcal{O}(1/\epsilon)$. However, a naive implementation of amplitude amplification ends up with some issues. One straightforward way to employ this algorithm is to measure individual local terms in the Hamiltonian (after decomposing them further into a linear combination of Pauli operators). However, the issue is that there are $\Omega(L^2)$ terms in the Hamiltonian. If we were to measure them altogether, it would incur a cost that scales quadratically with $L$, a substantial increase compared to the logarithmic scaling in $L$ in Section~\ref{sec:resource_estimate_incoherent}. If the state prepared by the DMERA circuit is translationally invariant, one could simply pick one of the sites and measure the local term acting nontrivially on that site. However, in practice it is difficult to ensure the translational invariance of the state. Yet another alternative is to sample uniformly randomly over the sites [Eq.~\eqref{eq:sampling_over_sites}], each time employing the amplitude estimation algorithm for the randomly chosen observable. Unfortunately, in this case, we would again restore the $\Omega(1/\epsilon^2)$ scaling because this is a sampling-based approach. Thus these naive applications of amplitude estimation algorithm are undesirable.

Motivated by these observations, we propose a novel method that retains the $\mathcal{O}(1/\epsilon)$ scaling (where $\epsilon$ is the error in the energy per site) while using at most logarithmic number of qubits (in $L$). We remark that prior works have already advocated using variants/improvements of the amplitude estimation algorithm~\cite{Brassard2002} for variational energy calculations~\cite{Wang2019,Wang2021,Zhang2022,Katabarwa2021}. However, it is a priori not obvious if these approaches can be used to achieve the desired $\mathcal{O}(1/\epsilon)$ accuracy using at most logarithmic number of qubits (in $L$), because the original DMERA circuit [Section~\ref{section:dmera}] is defined on $L^d$ qubits.

\begin{figure*}[t]
  \includegraphics[width=1.0\textwidth]{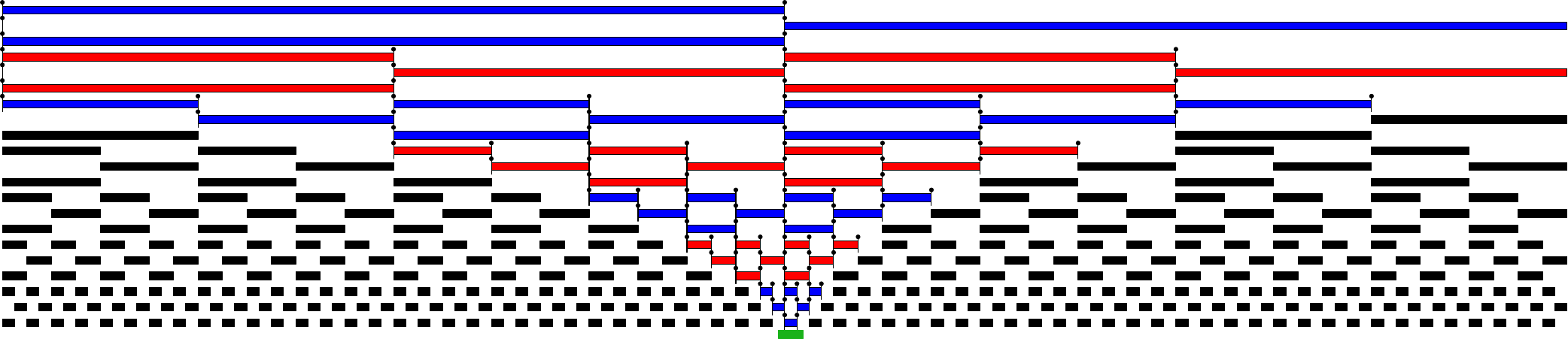}
  
  \includegraphics[width=1.0\textwidth]{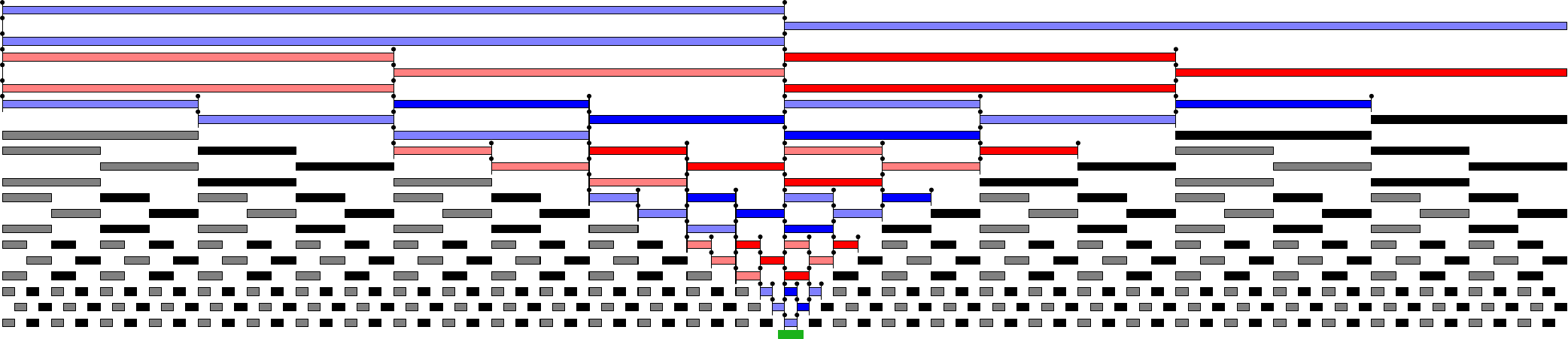}
  
  \includegraphics[width=1.0\textwidth]{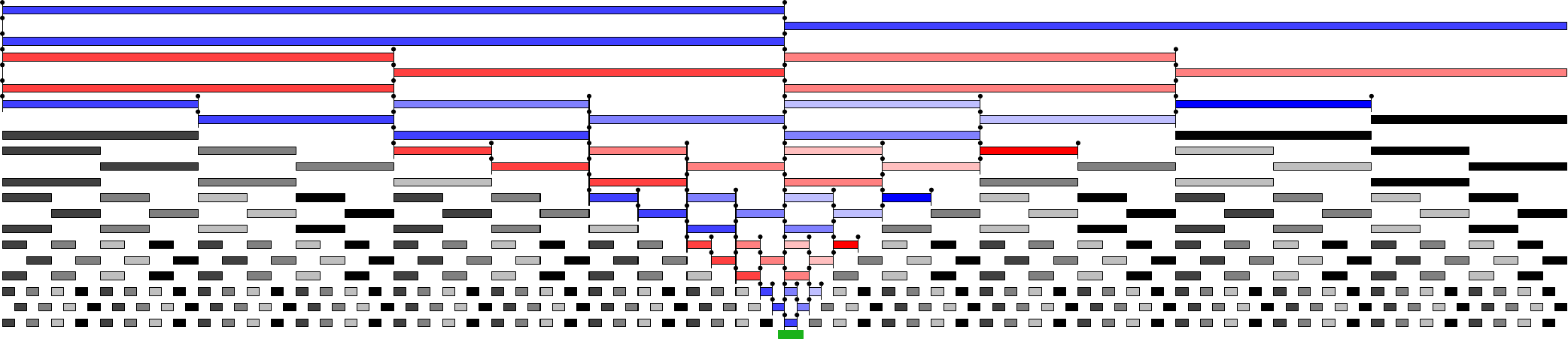}
  
  \caption{DMERA with a with translationally invariant gates with period $2$ (top), $4$ (middle) and $8$ (bottom). Here we chose $n=7$ and $D=3$ for concreteness. Within each layer, the gates with the same level of gradation are chosen to be the same gates. Blue/red gates are the gates that appear in the past causal cone of a nearest-neighbor observable (green). \label{fig:periodic_dmera}}
\end{figure*}

Without going into the details of our approach, let us first provide an executive summary on its complexity. An important parameter that determines the complexity is the \emph{period} of the DMERA circuit within each scale. For example, for $r=2$, within each scale, if all the gates are chosen identically, shifting all the qubits by $2$ (in the unit of the underlying lattice) yields the identical circuit. In this case, the period is $2$. More generally, the period can be increased by (i) choosing a larger $r$ or (ii) choosing the gates in a non-uniform way [Fig.~\ref{fig:periodic_dmera}]. Thus the period depends on two parameters: (i) the quantum circuit architecture and (ii) the pattern of the gates being chosen. 

Throughout this Section, for simplicity we shall assume the same period for different directions, and denote the period as $\mathcal{T}$. The overall complexity of the algorithm can be summarized as follows. 
\begin{itemize}
\item Gate count: $\mathcal{O}\left(\frac{ \max(D,\mathcal{T})^{d+1} \log  N}{\epsilon}\right)$
\item Number of qubits: $\Theta(\max (D, \mathcal{T})^d \log N)$.
\end{itemize}
Recall that $\mathcal{T}$ is a period, $N = 2^{nd}$ is the system size, $D$ is the depth of the DMERA per unit layer, and  $\epsilon$ is the relative precision in the energy. Thus, provided that the period $\mathcal{T}$ is sufficiently small compared to $D$, the extra cost in the gate count is negligible. Number of qubits required is increased, but only slightly, by a factor logarithmic in $N$.

We remark that $\mathcal{T}$ being a small number is a reasonable assumption. In many existing examples, the gates are chosen in a translationally invariant way~\cite{Evenbly2010,Olund2020,Miao2021,Miao2023a}, which immediately implies $\mathcal{T}=\mathcal{O}(1)$. Therefore, under such assumption one can achieve a gate count of $\mathcal{O}(1/\epsilon)$ and a qubit count of at most $\mathcal{O}(\log N)$. 

The rest of this Section is organized as follows. We first explain in Section~\ref{sec:amplitude_estimation_alg} the amplitude estimation algorithm~\cite{Brassard2002}, which is the algorithm that our algorithm builds upon. We consider a few ways in which this algorithm can be applied to our setup in a ``black box'' fashion, and explain why these approaches are inadequate for our purpose. In Section~\ref{sec:exploiting_dmera}, we emphasize a special structure of DMERA which we can utilize to obtain the desired complexity.

\subsection{Amplitude Estimation algorithm}
\label{sec:amplitude_estimation_alg}

In the amplitude estimation algorithm~\cite{Brassard2002}, one assumes access to a unitary $U$ that prepares some state $|\psi\rangle$ from a simple product state,. e.g., $|0\ldots 0\rangle$ and a reflection $R=I - 2P$, where $P$ is a projection. The cost of this algorithm can be quantified in terms of the number of invocation of the Grover-type operator:
\begin{equation}
  \mathcal{Q} = U R_0 U^{\dagger}R,\label{eq:qae_operator}
\end{equation}
where $R$ is the reflection about the $|0\ldots 0\rangle$ state. This algorithm estimates  $\langle \psi | P |\psi\rangle$ with an additive precision of $\epsilon$, using $\mathcal{O}(1/\epsilon)$ invocations of $\mathcal{Q}$.

It is well-known that the amplitude estimation algorithm can be used to estimate expectation values of certain observables with $\mathcal{O}(1/\epsilon)$ complexity, assuming one has access to a gate of the form of $e^{i\theta O}$ for an observable $O$ that one intends to measure and a set of $\theta\in \mathbb{R}$. Let us remark that if the observable of interest can be decomposed into a linear combination of $\mathcal{O}(1)$ number of multi-qubit Pauli matrices, the fact that this is possible is rather obvious; in this case one can simply choose a reflection as one of the individual Paulis, estimate their expectation values individually, and take a sum.

As we discussed before, a naive application of the amplitude estimation algorithm to DMERA yields the $\mathcal{O}(1/\epsilon)$ complexity, but with caveats. Doing so appears to increase the gate complexity in $N$ (system size) from $\mathcal{O}(\log N)$ to $\mathcal{O}(N\log N)$. Randomly selecting an observable (in the spirit of Section~\ref{sec:resource_estimate_incoherent}) and applying amplitude estimation for the set of chosen observables yields a $\mathcal{O}(\log N)$ scaling, but now the scaling in $\epsilon$ becomes $\mathcal{O}(1/\epsilon^2)$, unless the variance of the expectation value of the observable (with respect to the choice of the observable happens) is zero. This is unlikely, because even if the gates in the DMERA are chosen in a translationally invariant way, the resulting state may not be translationally invariant.

\subsection{Coherent Sampling: Amplitude Estimation for DMERA}
\label{sec:exploiting_dmera}

The naive approaches explained above fail to achieve $\mathcal{O}(\epsilon^{-1}\log N)$ gate complexity. Here we propose a method that achieves this goal, referred to as the \emph{coherent sampling} approach. The key idea is to construct a compact circuit that prepares a purification of the uniform mixture of the reduced density matrices that support individual terms of the Hamiltonian. Suppose, for example, that we are interested in estimating the expectation value of a nearest-neighbor Pauli operator $P$ supported on $(x,y)$ and $(x+1, y)$, averaged uniformly over $x$, which we denote as $\bar{P}$. To estimate the expectation value of this observable, we will seek to prepare a purification of the reduced density matrix over $(x,y)$ and $(x+1, y)$, averaged uniformly over both $x$ and $y$. It is then easy to see that the the expectation value of $\bar{P}$ with respect to the original state is equal to the expectation value of $P$ with respect to the averaged density matrix. Thus, if we can construct a circuit that prepares a purification of the averaged reduced density matrix, we can apply the amplitude estimation algorithm to that circuit to measure expectation value of $\bar{P}$ with $\mathcal{O}(1/\epsilon)$ complexity.

Such a purification can be constructed by noting that within every scale of the DMERA, there are at most $\mathcal{T}^d$ distinct circuits, all related to each other by shifting the qubits in the $x$- and $y$-directions. As a simple example, consider a one-dimensional example, shown in Fig.~\ref{fig:pcc_regularity}. Here one can see that the past causal cone within a single scale, for all local observables, can be obtained by shifting a sufficiently wide past causal cone and possibly restricting it to a subset of qubits. Therefore, one can obtain the requisite averaged reduced density matrix by applying a random shift as we transition from one scale to another. Importantly, this can be done coherently, as we explain below.

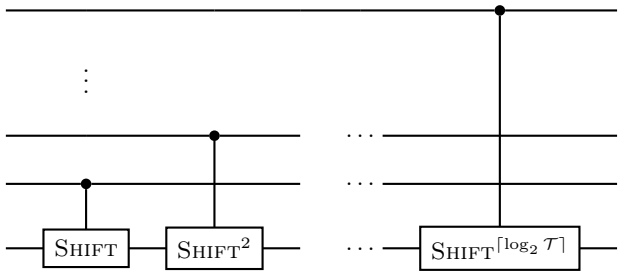
\begin{figure}[h]
  \begin{quantikz}
    \qw & \qw & \qw & \qw & \qw & \ctrl{4} &\qw \\
     & \vdots & & & & &\\
    \qw & \qw & \ctrl{2} &\qw & \ \ldots \ &\qw &\qw \\
   \qw & \ctrl{1} & \qw & \qw & \ \ldots \ &\qw &\qw \\
    \qwbundle{n} & \gate{\textsc{Shift}}& \gate{\textsc{Shift}^2} &\qw & \ \ldots \ & \gate{\textsc{Shift}^{\lceil \log_2 \mathcal{T}\rceil}} & \qw
  \end{quantikz}
  \caption{Using c\textsc{Shift} to randomize the connection between different scales.\label{fig:controlled-shift}}
\end{figure}

A useful subroutine is \textsc{Shift}. This operation shifts the qubits in the past causal cone of the DMERA, either in the $x$- or $y$-direction. Let us denote the linear size of the past causal cone as $\ell$. Specifically, let $\textsc{Shift}$ be a unitary defined as
\begin{equation}
    \textsc{Shift}:  \bigotimes_{k=0}^{\ell-1}|\varphi_k\rangle_k \to  \bigotimes_{k=0}^{\ell-1}|\varphi_{\sigma_1(k)}\rangle_k,
\end{equation}
where $\sigma_1$ is the cyclic permutation, i.e., $\sigma_1(0) =1, \sigma_1(1)=2, \ldots \sigma_1(\ell-1)=0$ and $|\varphi_k\rangle_k$ is a $\ell$-qubit state. We can similarly define c\textsc{Shift} (an acronym for controlled-\textsc{Shift}) as
\begin{equation}
  c\textsc{Shift}: |x\rangle \bigotimes_{k=0}^{\ell-1}|\varphi_k\rangle_k \to |x\rangle \bigotimes_{y=0}^{\ell-1}|\varphi_{\sigma_x(k)}\rangle_k,
\end{equation}
where $\sigma_0$ is the identity and $\sigma_x = (\sigma_1)^x$ is a cyclic permutation by $x \in \{0, \ldots, \ell-1\}$. For an implementation of $c\textsc{Shift}$, see Fig.~\ref{fig:controlled-shift}.

We note that any cyclic permutation over $m$ elements can be decomposed into at most $m$ \textsc{SWAP}s. In two dimensions, because there are at most $\ell$ rows/columns in the past causal cone, any power of \textsc{Shift} can be realized using at most $\ell^2$ \textsc{SWAP}s. Therefore, using the construction in Fig.~\ref{fig:controlled-shift}, we conclude the cost of $c\textsc{Shift}$ to be at most $\ell^2 \lceil \log_2 \mathcal{T} \rceil$ controlled-\textsc{SWAP}s. Each controlled-\textsc{SWAP} can be decomposed into one \textsc{Toffoli} and two \textsc{CNOT} gates. Thus, c\textsc{Shift} costs at most $\ell^2 \lceil \log_2 \mathcal{T} \rceil$ \text{Toffoli}s and $2\ell^2 \lceil \log_2 \mathcal{T} \rceil$ \textsc{CNOT} gates. In particular, it consumes at moat $8\ell^2 \lceil \log_2 \mathcal{T} \rceil$ $T$-gates~\cite{Gidney2018}.

Now let us discuss how to use these operations to obtain the averaged density matrix in DMERA. One can do this by essentially randomizing the way in which the output from one scale is connected to the input to the other scale. In 1D DMERA, for each scale, there are exactly $\ell$ possible ways in which the output from one scale is connected to the input to the next (lower) scale. The number of ancilla qubits needed is $\lceil \log_2 \mathcal{T} \rceil$ using the standard circuit construction. Similarly, in 2D DMERA, for each scale, there are exactly $\ell^2$ possible ways in which the output from one scale is connected to the input to the next (lower) scale. The total ancilla qubit number is $2\lceil \log_2 \mathcal{T} \rceil$ and the gate count being $2\ell^2 \lceil \log_2 \mathcal{T} \rceil$ controlled-\textsc{SWAP}s.

To coherently randomize these shifts, we must prepare the ancillary register in the uniform superposition over $\{0, \ldots, \mathcal{T}-1\} \times \{0, \ldots, \mathcal{T}-1 \}$. The cost of this is known to be $\mathcal{O}(\log \mathcal{T})$; see Ref.~\cite{Babbush2018} for details. 
 
\subsection{Resource estimate}
\label{sec:resource_estimate_coherent}

In this Section, we perform a more detailed resource estimate of the coherent sampling algorithm. In particular, we focus on estimating the overall $T$-gate cost and the number of qubits associated to the implementation of $\mathcal{Q}$ [Eq.~\eqref{eq:qae_operator}]. Implementation of $\mathcal{Q}$ consists of three components: \textsc{cShift}, preparation of the uniform superposition state over $\{0,\ldots, \mathcal{T}-1 \}\times\mathcal{T}-1 \}$, and application of the gates lying in the past causal cone.

Here is a breakdown of the $T$-gate count. The \textsc{cShift} and the uniform superposition state preparation incurs $\mathcal{O}(\ell^d \log \mathcal{T})$ and $\mathcal{O}(d\log \mathcal{T}$) respectively, as discussed in Section~\ref{sec:exploiting_dmera}. The number of two-qubit gates lying in the past causal cone can be estimated to be $\mathcal{O}(\ell^{d+1}\log N)$, where $\ell = \Theta(\max(D, \mathcal{T}))$ [Section~\ref{sec:pcc},~\ref{sec:exploiting_dmera}]. Thus the overall cost is $\mathcal{O}(\max(D,\mathcal{T})^{d+1} \log N)$. Recalling that the query complexity(for $\mathcal{Q}$) of the amplitude estimation algorithm~\cite{Brassard2002} is $\mathcal{O}(1/\epsilon)$,we arrive at the overall cost of $\mathcal{O}\left(\max(D,\mathcal{T})^{d+1} \epsilon^{-1} \log N\right)$.\footnote{There are additional gates used in the amplitude estimation algorithm, such as the implementation of the reflection $R$. However, the associated cost is negligible in comparison.}

The number of qubits can be estimated as follows. For implementing the gates in the past causal cone, there are $\Theta(\max(D, \mathcal{T})^d \log N)$ qubits. For the \textsc{cShift} and the preparation of uniform superposition state, there are $\Theta(\log \mathcal{T})$ additional qubits needed. Thus the overall qubit count is $\Theta(\max(D, \mathcal{T})^d \log N)$.

\section{Conclusion}
\label{sec:conclusion}

In this paper, we explored the possibility of using variational ansatzes such as DMERA~\cite{Kim2017} to study the ground state properties of a system consisting of a large number of fermions interacting with each other. The main strength of this approach is that it requires a relatively smaller number of qubits compared to the approaches based on QPE. For the specific case we studied, consisting of a system of size $64\times 64$, we observed that indeed a large reduction in the number of qubits is possible [Section~\ref{sec:resource_estimate_incoherent}].

We also studied the pertinent issues concerning variational approaches, such as the large number of measurements needed~\cite{Wecker2015} and the Barren plateau problem~\cite{McClean2018}. We found that the measurement problem still persisted even for a simple model (e.g., Fermi-Hubbard model in two spatial dimensions), suggesting that the conventional QPE-based approaches will be more advantageous if a sufficiently large quantum computer is available. The root of this problem lies in the inevitable $\mathcal{O}(1/\epsilon^2)$ scaling for the precision $\epsilon$ that is present in any sampling-based approach. Fortunately, we found that it is possible to restore the Heisenberg scaling whilst only incurring an additional logarithmic overhead in the system size [Section~\ref{sec:amplitude_estimation_alg}]. 

For the Barren plateau problem, we found that the typical magnitude of the gradient depends greatly on the location of the gate; the deeper it is in the circuit, smaller the gradient is. More precisely, we can organize the DMERA circuit into $n=\log_2 N$ layers, each corresponding to a different length scale [Fig.~\ref{fig:pcc_regularity}]. The gates appearing the coarsest length scale have a variance decaying exponentially in $Dn$, yielding a polynomially small number. However, at the finest length scale, the variance decays exponentially in $D$ alone, which is a constant indepent of the system size. At the $s$'th layer (setting $s=0$ to be the finest and $s=n$ to be the coarsest scale), the variance decays exponentially with $ns$. Thus for DMERA, the Barren plateau problem can be avoided as a function of system size.

There are several natural questions to pursue, which we leave for future work. First, a more detailed comparison between the coherent and incoherent sampling approach --- taking into account of the constant factors --- would be desirable. Second, it will be desirable to develop a theory on how well the approximation error scales with $D$. The existing studies~\cite{Evenbly2016,Haegeman2017} are limited to non-interacting systems. Would this picture change or remain intact in the presence of interaction? Perhaps experiments involving small-scale quantum computers will be illuminating. Third, understanding the expressiveness of the 2D DMERA remains as an important open problem. In Ref.~\cite{Haghshenas2022}, the authors investigated how well tensor network ansatzes suitable for studying many-body systems in one spatial dimension~\cite{Kim2017,Kim2017a,Foss-Feig2021} can approximate ground states of physical interest.\footnote{We remark, however, that Ref.~\cite{Haghshenas2022} also studies how well certain two-dimensional systems are approximated by a one-dimensional ansatz such as MPS.} Extending such a study to two spatial dimensions is an important open problem. Lastly, due to a very modest resource requirement reported in Section~\ref{sec:resource_estimate_incoherent}, the fault-tolerance overhead needed to make the effect of noise negligible may be much more benign than the requirements for other approaches based on QPE. Quantifying this amount will be another interesting direction to pursue. 

\section*{Acknowledgement}
This work was supported in part by funding from Lockheed Martin.

\bibliography{bib}

\appendix

\section{Compilation approaches}
\label{appendix:space_efficient_compilation}

In this Appendix, we introduce compilation algorithms used in obtaining Fig.~\ref{fig:dmera_incoherent_cost}. Specifically, we introduce two algorithms, referred to as Algorithm~\ref{alg:algo_a} and Algorithm~\ref{alg:algo_b}. Algorithm~\ref{alg:algo_a} has the advantage of being relatively fast, but leads to suboptimal number of qubits. In contrast, Algorithm~\ref{alg:algo_b} leads to a reduced qubit cost but is slower than Algorithm~\ref{alg:algo_a}. 

Throughout this Appendix, we shall assume that a circuit is described by the following data structure. A \texttt{gate} is an object with the property called \texttt{support}, which specifies the set of qubits the \texttt{gate} acts on. A \texttt{layer} is a collection of \texttt{gate}s, each of which have disjoint \texttt{support}s. Intuitively, a layer can be thought as a unit-depth quantum circuit. A \texttt{circuit} is a sequence of \texttt{layer}s. To summarize, a \texttt{circuit} is a sequence of \texttt{layer}s, each of which consists of \texttt{gate}s that commute with each other.

The goal of the compilation algorithms is to assign the qubits appearing in the \texttt{support} of each \texttt{gate} to the set of qubits in the quantum computer. The algorithms discussed below achieve this goal, with additional gates that reset the state of the qubit. The cost of these extra gates are negligible and they are useful for reducing the number of qubits. 

We shall assume access to the following simple functions, whose implementation is straightforward. \texttt{Qubits} is a function that, applied to an object, returns a set of qubits that the object nontrivially acts on. for instance, applied to a \texttt{layer} object, it returns the union of the \texttt{support}s of the \texttt{gate}s in the \texttt{layer}. Applied to a \texttt{circuit} layer, it returns the union of \texttt{Qubits(layer)} for all the \texttt{layer}s within the \texttt{circuit}. \texttt{Length} is a function that, applied to an object discussed above, returns the number of elements of the object. For instance, applied to a \texttt{circuit} object, this function returns the number of \texttt{layer}s within the \texttt{circuit} object. Applied to a \texttt{layer}, it returns the number of \texttt{gate} objects within the \texttt{layer}. Applied to the set of qubits, it returns the cardinality of the set.

The basic idea behind Algorithm~\ref{alg:algo_a} is to iterate through each layer in the following way. Given a layer, we can divide the circuit into two pieces. One of them is the ``past,'' which we denote as \texttt{Past(layer, circuit)}. This is a sequence of \texttt{layer}s within the \texttt{circuit} up to the \texttt{layer} under consideration. The other is the ``future,'' which we denote as \texttt{Future(layer, circuit)}. This is the sequence of \texttt{layer}s appearing after the \texttt{layer} under consideration, within the \texttt{circuit}. Note that the qubits in the past not appearing in the future need not be kept, because they are never used afterwards. Therefore, we can simply reset these qubits, freeing them up for future use; this corresponds to Line 8 of Algorithm~\ref{alg:algo_a}.  

The pseudocode description of Algorithm~\ref{alg:algo_a} uses two additional operations, denoted as \texttt{Assign} and \texttt{Remove}. Both these operations implicitly assume that there is a globally accessible lookup table from the set of qubits appearing in the circuit description to the set of qubits in the quantum computer. The set of qubits available on a quantum computer is assumed to be an ordered list, and the \texttt{Assign} makes/updates the assignment in the following greedy way. If the assignment for a qubit in the circuit description is already specified in the lookup table, \texttt{Assign} simply applies this lookup table to the \texttt{support}s of the \texttt{gate}s within each \texttt{layer}. If not, \texttt{Assign} updates the assignment to the first qubit (on the quantum computer) that is not yet assigned and then repeat the aforementioned procedure. The \texttt{Remove} operation removes the assignment in the lookup table for the specified set of qubitss (in the circuit description).

\begin{algorithm}[H]
  \caption{High-level description of Algorithm~\ref{alg:algo_a}.} \label{alg:algo_a}
  \hspace*{\algorithmicindent} \textbf{Input:} \texttt{Circ} \\
  \hspace*{\algorithmicindent} \textbf{Output:} \texttt{CircNew}
  \begin{algorithmic}[1]
    \State $\ell \gets \texttt{Length(Circ)}$
    \State \texttt{CircNew} $\gets$ \texttt{[]}
    \For{$i$ \texttt{in range}$(\ell)$}
    \State \texttt{CircNew.append(Assign(Circ[i]))}
    \State \texttt{PastCirc} $\gets$ \texttt{Circ[:i]}
    \State \texttt{FutureCirc} $\gets$ \texttt{Circ[i:]}
    \State \texttt{RemoveQ} $\gets$ \texttt{Qubits(FutureCirc)-Qubits(PastCirc)}
    \State \texttt{CircNew.append(Reset(Assign(RemoveQ)))}
    \State \texttt{Remove(RemoveQ)}
    \EndFor
    \Return \texttt{CircNew}
  \end{algorithmic}
\end{algorithm}

To explain the basic idea behind Algorithm~\ref{alg:algo_b} it will be useful to first start with a motivating observation. Without loss of generality, let \texttt{q} be a qubit appearing in the circuit and let \texttt{PCC(q, Circ)} be the past causal cone of \texttt{q} with respect to the \texttt{circuit} \texttt{Circ}. For any gate in \texttt{Circ} not appearing in the past causal cone, it is possible to move that gate past all the gates appearing in the past causal cone. The proof of this claim is straightforward. If the gate appears after the past causal cone, the claim is true by construction. Otherwise, if a gate lies between gates appearing in the past causal cone and does not commute with at least one gate in the past causal cone appearing afterwards, that gate must necessarily lie in the past causal cone; this is a contradiction. Therefore, for any \texttt{q}, the unitary generated by \texttt{Circ} can be decomposed into the unitary generated by \texttt{PCC(q, Circ)} followed by other gates.

The basic idea of Algorithm~\ref{alg:algo_b} is to compile \texttt{Circ} by decomposing it into a past causal cone of a qubit for which the number of qubits in the past causal cone is minimized, followed by the remaining set of gates. The main virtue of this construction is that, once the past causal cone of the qubit is applied, that qubit can be removed, akin to the process described in Algorithm~\ref{alg:algo_a}. Then the same procedure can be applied to the remainder of the circuit repeatedly. 

The pseudocode description of Algorithm~\ref{alg:algo_b} makes use of the operations used in Algorithm~\ref{alg:algo_a} such as \texttt{Assign} and \texttt{Remove}. However, Line 4 and 7 requires more explanation. In Line 4, \texttt{PCCSizes} is a dictionary data structure with the keys corresponding to the set of qubits in the circuit and the values being the number of qubits in the past causal cone of that qubit. In Line 7, the subtraction of \texttt{PCC(Qmin, Circ)} from \texttt{Circ} means simply removing the gates in \texttt{PCC(Qmin, Circ)} from \texttt{Circ}. To be more precise, view \texttt{Circ} as a sequence of \texttt{gate}s, each with a unique identification number.\footnote{For instance, even if the same gate is used twice, those two different instantiations of the same gate will be associated with two different identification numbers.} Line 7 means removing the set of gates --- labeled in terms of their identification number --- in the past causal cone from \texttt{Circ}.

\begin{algorithm}[H]
  \caption{High-level description of Algorithm~\ref{alg:algo_b}. } \label{alg:algo_b}
  \hspace*{\algorithmicindent} \textbf{Input:} \texttt{Circ} \\
  \hspace*{\algorithmicindent} \textbf{Output:} \texttt{CircNew}
  \begin{algorithmic}[1]
    \State \texttt{CircNew} $\gets$ \texttt{[]}
    \While{\texttt{Circ} is non-empty }
    \State \texttt{Qs} $\gets$ \texttt{Qubits(Circ)}
    \State \texttt{PCCSizes} $\gets$ \texttt{\{ Q:[Length(Qubits(PCC(Q, Circ))) for Q in Qs]\}}
    \State \texttt{Qmin} $\gets$ \texttt{argmin(PCCSizes)}
    \State \texttt{CircNew.append(Assign(PCC(Qmin, Circ)))}
    \State \texttt{Circ} $\gets$ \texttt{Circ - PCC(Qmin, Circ)}
    \State \texttt{CircNew.append(Reset(Assign(Qmin)))}
    \State \texttt{Remove(Qmin)}
    \EndWhile
    \Return \texttt{CircNew}
  \end{algorithmic}
\end{algorithm}

Let us make a side remark: that in practice Algorithm~\ref{alg:algo_b} is substantially slower than Algorithm~\ref{alg:algo_a}. Due to this reason, in making Fig.~\ref{fig:dmera_incoherent_cost} we simply estimated the number of qubits needed on a quantum computer, instead of running the algorithm in Algorithm~\ref{alg:algo_b}.\footnote{The main bottleneck in practice is Line 7 of Algorithm~\ref{alg:algo_b}. A judiciously choice of data structure seems to be necessary to make this approach practical.}

Now that we have explained the two algorithms, it is interesting to discuss which of them is of more practical use. As explained above, Algorithm~\ref{alg:algo_a} is faster but leads to a suboptimal number of qubits. On the other hand, Algorithm~\ref{alg:algo_b} is slower but leads to a smaller number of qubits. We believe, at least in the foreseeable future, it is desirable to use Algorithm~\ref{alg:algo_b}. Qubits, especially logical qubits, are likely to remain a scarce resource, and as such, it is always desirable to minimize that number. Moreover, the compilation can be done offline prior to the quantum computation. Since classical computation resources are significantly cheaper than the quantum ones, a large compilation cost appears to be an acceptable cost.

\section{Compiling Fermionic DMERA}
\label{appendix:fermions}
In this Appendix, we introduce a method to compile a fermionic variant of DMERA, fDMERA for short. This variant of DMERA has the same circuit architecture as DMERA, but assumes that the global Hilbert space and the gates used are fermionic. Fermionic gate refers to a gate acting on a set of (second-quantized) fermions which preserve the fermion parity. A universal set of fermionic gates was first introduced by Bravyi and Kitaev~\cite{Bravyi2002}, summarized below:
\begin{equation}
  \begin{aligned}
    &\exp(\theta (e^{i\varphi}a_i^{\dagger}a_j + h.c.  )), \\
    &\exp(\theta (e^{i\varphi} a_i^{\dagger} a_j^{\dagger} + h.c.)), \\
    &\exp(\theta n_i n_j),
  \end{aligned}
  \label{eq:fermionic_gates}
\end{equation}
where $\theta, \varphi \in \mathbb{R}$ are real parameters, $a_i^{\dagger}$ and $a_i$ are fermion creation and annihilation operators respectively, and $n_i = a_i^{\dagger} a_i$. Fermionic DMERA can be simply thought as a quantum circuit with a hierarchical structure (see Section~\ref{section:dmera}), consisting of fermionic gates. These circuits was studied both numerically~\cite{Corboz2010} and analytically~\cite{Haegeman2017}. 

Since quantum computers that are likely going to be available will consist of qubits, to use fermionic ansatzes a conversion from a fermionic circuit to a circuit acting on a set of qubits is needed. One could, in principle, use the standard Jordan-Wigner(JW) transformation~\cite{JW1928}. However, JW transformation maps few-body fermionic observables to a many-body spin observable, which is undesirable. There are other alternative approaches which preserve locality~\cite{Verstraete2005,Whitfield2016,Steudtner2019,Bravyi2002,Setia2019,Jiang2019,Derby2021,Chen2022}. Upon applying these methods, one would be able to convert the gates in Eq.~\eqref{eq:fermionic_gates} to unitaries acting on $\mathcal{O}(1)$ qubits. However, such a mapping often increases the number of qubits needed.

Motivated from these facts, we provide an alternative approach that can minimize the number of qubits needed. The basic idea is to first compute the past causal cone \emph{without converting the fermions to qubits}. This ends up being a fermionic circuit acting on a smaller set of fermions. By applying the JW transformation to this reduced circuit, we can convert this circuit to a qubit circuit without incurring any extra qubit overhead. 

Recall that in Appendix~\ref{appendix:space_efficient_compilation} we introduced methods to reduce the number of qubits needed, by keeping track of qubits that are no longer in use and resetting them, freeing them up. Analogous tricks can be applied to fermions as well. However, what is not so clear is what it means to reset a fermion when we convert that process into a physical operation acting on a set of qubits, via the JW transformation. This is what we intend to discuss primarily in this Appendix.

Without loss of generality, suppose we would like to reset the $k$'th fermion to be in the unoccupied state. In the fermionic occupation basis, the operation that we intend to perform is
\begin{equation}
  |x\rangle_f \to |x_k\rangle_f,
\end{equation}
where $|x\rangle_f$ is a basis state of the Fock basis represented by a bit-string $x$ and $x_k$ can be obtained by setting the $k$'th bit of $x$ to $0$. This can be done by (i) measuring the $k$'th fermion occupation number of and (ii) if the occupation number is $1$, apply the annihilation operator $a_k$.

Under the JW transformation, the occupation number $n_k$ is related to the Pauli operators acting on qubits as $Z_k = 2n_k - I$. The annihilation operator $a_k$ in terms of the Pauli operators is
\begin{equation}
  \left( \bigotimes_{j=1}^{k-1} Z_k \right) \bigotimes \frac{X_k - iY_k}{2},
\end{equation}
whose action on the state in which the $k$'th qubit is set to $Z_k = +1$ is equivalent to the action of the following operator
\begin{equation}
    \left( \bigotimes_{j=1}^{k-1} Z_k \right) \bigotimes X_k.
\end{equation}
Thus, to reset the $k$'th fermion to the unoccupied state, it suffices to measure $Z_k$ and apply$\left( \bigotimes_{j=1}^{k-1} Z_k \right) \bigotimes X_k$ only if $Z_k=+1$. Resetting the fermion to the occupied state can be performed in a similar way, by instead using the JW transformation of the fermionic creation operator. 

\end{document}